\documentclass[final,11pt]{article}
\usepackage[margin=1in,letterpaper]{geometry}
\usepackage{amsmath,amssymb,amsfonts}
\usepackage[pdftex]{graphicx}
\usepackage{multirow} 
\usepackage{xcolor-solarized}
\usepackage{booktabs}
\usepackage{setspace}
\usepackage{authblk}
\usepackage[mathlines]{lineno}
\usepackage[bf,sf,compact]{titlesec}
\usepackage{natbib}
\usepackage{xr}
\usepackage{pdfpages}
\providecommand{\abs}[1]{\left\lvert#1\right\rvert}
\providecommand{\en}{\ensuremath{\epsilon_{\rm n}}}
\providecommand{\et}{\ensuremath{\epsilon_{\rm t}}}
\providecommand{\etr}{\ensuremath{\epsilon_{\rm tc}}}
\providecommand{\etht}{\ensuremath{\hat{\epsilon}_{\rm t}}}
\providecommand{\I}{\ensuremath{I_{\rm /G}}}
\providecommand{\tr}{\ensuremath{^{\rm T}}}
\renewcommand{\vec}[1]{\ensuremath{\boldsymbol {#1}}}

\def\fig{Fig.}
\def\eq{equation}
\def\eqs{equations}
\def\sec{section}
\def\mc{Monte Carlo~}
\def\supp{supplement}

\begin{document}
\title{How to run on rough terrains} 
\author[1,2]{Nihav Dhawale}
\author[3]{Shreyas Mandre}
\author[1]{Madhusudhan Venkadesan\thanks{Address correspondence to {\ttfamily mv@classicalmechanic.net}}}
\affil[1]{\footnotesize Department of Mechanical Engineering and Materials Science, Yale University, New Haven, CT 06520}
\affil[2]{\footnotesize National Centre for Biological Sciences--Tata Institute of Fundamental Research, Bangalore, Karnataka 560065}
\affil[3]{\footnotesize School of Engineering, Brown University, Providence, RI 02912}
\date{}
\maketitle

\vspace{-0.2in}
\section*{Abstract}  
Stability of running on rough terrain depends on the propagation of perturbations due to the ground.
We consider stability within the sagittal plane and model the dynamics of running as a two-dimensional body with an alternating aerial and stance phase.
Stance is modeled as a passive, impulsive collision followed by an active, impulsive push-off that compensates for collisional losses.
Such a runner has infinitely many strategies to maintain periodic gaits on flat ground.
However, these strategies differ in how perturbations due to terrain unevenness are propagated.
Instabilities manifest as tumbling (orientational instability) or failing to maintain a steady speed (translational instability).
We find that open-loop strategies that avoid sensory feedback are sufficient to maintain stability on step-like terrains with piecewise flat surfaces that randomly vary in height.
However, these open-loop runners lose orientational stability on rough terrains whose slope and height vary randomly.
Only by avoiding tangential collisions is orientational stability recovered.
Tangential collisions may be avoided through leg-retraction to match foot and ground speed at touch down.
By analyzing the propagation of perturbations, we derive a single dimensionless parameter that governs stability and guides the design and control of both biological and robotic runners.

\clearpage
\section{Introduction} 
Legged terrestrial animals run stably on rough terrains, despite potential difficulties such as sensory latencies and the highly dynamic nature of running.
Our current understanding of how running animals negotiate rough terrains is based on studies where the animal experiences obstacles in the form of a single step up or down \citep{Daley2006,Birn2012}, or a random sequence of up and down steps \citep{Grimmer2008,Voloshina2015}.
However, natural terrains exhibit not only variations in height but also in slope, and it is unclear how our understanding of running on step-like terrains translates to such natural terrains.

Mathematical studies of running over rough terrains reflect the experiments and focus on stability when running on step-like terrains that are piecewise flat \citep{Daley2010,Blum2014,Karssen2015}.
Furthermore, models of runners with massless legs, such as the spring-legged-inverted-pendulum (SLIP) \citep{Blickhan1989,McMahon1990,Blickhan1993}, cannot distinguish between different slopes of the terrain and only respond to variations in height.
Assuming a massless leg enforces the ground reaction force vector to always align with the leg \citep{Srinivasan2008} regardless of the terrain's slope beneath the foot.
More detailed models that mimic the anatomy of specific animals or robots avoid this limitation of SLIP models \citep{Karssen2015}, at the cost of generalizability.
Thus there is a need for generalizable models of running that incorporate dependence on both terrain slope and height, and yet remain sufficiently abstract to glean principles that may underlie stability on rough terrains.

Stability may be governed by many factors, including sensory feedback control \citep{Pearson1993,Pearson1995,Dickinson2000}, the inherently stabilizing mechanical response of the animal's body \citep{Holmes2006}, energy dissipation within the body \citep{Daley2006}, and feed-forward strategies such as swing-leg retraction \citep{Seyfarth2003}.
The slowest is often sensory feedback control that has latencies comparable to or greater than the stance duration.
For example, at endurance running speeds for humans, the stance lasts around 200~ms \citep{Cavagna1964} and only slightly longer than the shortest proprioceptive feedback delay of 70--100~ms or visual feedback delay of 150--200~ms \citep{vanBeers2002}.
To better understand the inherent stability or instability of the dynamics of running, we consider only passive mechanical and anticipatory strategies in this study without relying on active feedback control.

Studies of running birds and the role of open-loop stability of running find that increased energy dissipation during stance may help stability when faced with an unexpected drop in terrain height \citep{Daley2006}.
Consistent with the role of energy dissipation, experiments with humans find that metabolic power increases by 5\% to run on step-like terrains versus flat ground \citep{Voloshina2015}.
Walking over rough terrains leads to an increase of 28\% in metabolic power \citep{Voloshina2013}, higher in both relative and absolute terms.
The difference in energetics may indicate that the dynamics of running are inherently less unstable, but such an analysis on natural rough terrains has not been carried out.
Therefore, we incorporate energy dissipation in our examination of open-loop strategies to address the effect of dissipation on stability.

Not relying on feedback control within a single stance does not preclude active strategies that rely on anticipation or internal models, sometimes called feed-forward strategies.
Computational studies of walking have demonstrated the role of look-ahead strategies that use the height and slope of the oncoming terrain in planning the control \citep{Byl2009}.
Evidence for the importance of feed-forward strategies for running come from computational studies of SLIP-like running dynamics \citep{Seyfarth2003} that show how swing-leg retraction automatically modulates the landing angle in response to unexpected variations in the terrain height.
However, these studies on running have not yet considered the effect of slope variations in the terrain.
Thus in our study, we analyze anticipatory strategies that incorporate the slope of the oncoming terrain.

An extreme and simplified approximation of running is that of a point mass with an impulsive and instantaneous stance followed by projectile flight.
Such an approximation appears as a natural solution to the problem of minimizing measures of metabolic energy consumption when the desired forward speed exceeds critical levels, and subject to other constraints such as step length \citep{Srinivasan2006a}.
SLIP-like models are an unfolding of these point-mass instantaneous-stance models to have finite stance duration.
They have helped us understand the kinetics of stance \citep{Holmes2006} and the energetics of producing forces \citep{Srinivasan2006a} on flat terrains, and the role of swing-leg retraction on piecewise flat terrains \citep{Blum2014}.
However, these point-mass models possess no sense of body orientation during the aerial phase and are therefore immune to falling by tumbling.

In this study, we unfold the point-mass, instantaneous-stance model by using a finite moment of inertia for the runner, while still maintaining an impulsive stance.
A finite moment of inertia defines a body orientation and thus enables the examination of the effect of angular momentum fluctuations induced by stance.
Such a model maps the net effect of the ground forces over stance as a linear impulse applied at the contact point and an angular impulse applied at the center of mass.
These impulses lead to a change in the linear and angular momentum of the whole runner because of the passive and active forces during stance.
As we discuss later, the angular impulse captures the effect of a finite stance duration and configuration changes during stance.

Section~\ref{sec:model} develops a sagittal-plane model that incorporates a finite moment of inertia, inelastic 2D collisions, and an active push-off so that both terrain slope and height variations affect stability.
In section~\ref{sec:mc}, we use Monte Carlo simulations with random variation of ground height and slope to examine open-loop strategies, the effects of energy dissipation, and strategies that anticipate the slope of the terrain.
Section~\ref{sec:stability} derives the linearized dynamical equations and analyzes their stability.
Using the linearization, we find a single dimensionless parameter that governs stability in \sec~\ref{sec:paramdepend}, which in turn guides morphological design for stability.
We conclude in section~\ref{sec:discussion} with a discussion of the limitations and generality of our analyses, its relationship to experimental results, and generate testable predictions for future experiments.

\section{Mathematical model of sagittal plane running}
\label{sec:model}

\begin{figure}
\centering
\includegraphics[width=\textwidth]{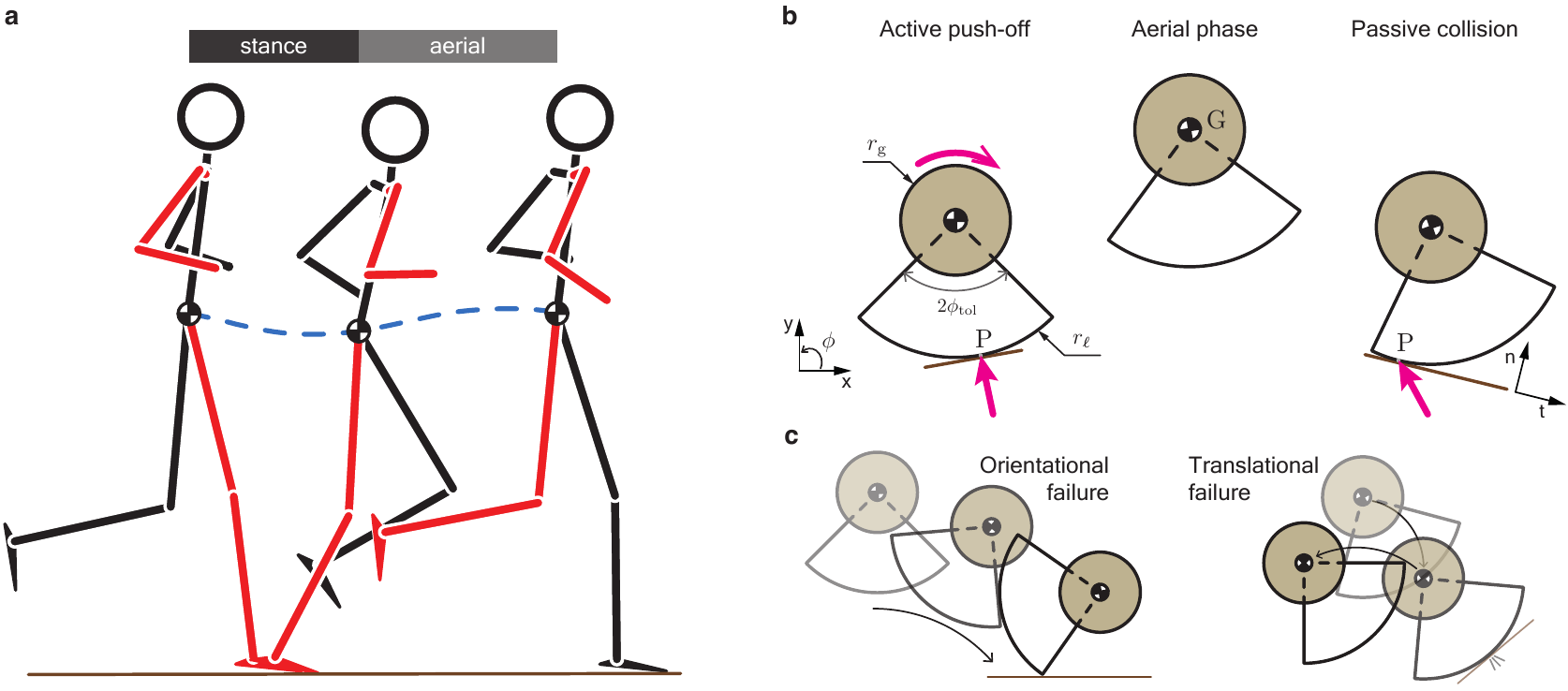} 
\caption{Bouncing as a model of running.
{\bfseries a}, The outline of a human running at 3.5~m/s, created from motion capture data, shows stance and aerial phases over a single step. The stance leg and ipsilateral arm are in red, and the center of mass trajectory is shown as a blue, dashed curve.
{\bfseries b}, The runner pushes-off the ground by applying a linear impulse $\vec{J}_{\rm imp}$ at the contact point P, and the effect of additional torques about the center of mass arising from configuration changes during stance are captured by an angular impulse $J_{\phi}$ at the center of mass.
At the end of the aerial phase, the runner undergoes a passive collision with the ground at the new contact point P. 
The momentum lost due to the collision in directions tangential and normal to the terrain surface is dictated by the parameters $\et$ and $\en$, respectively.
{\bfseries c}, The runner can fail in two ways: \emph{orientational failure} when orientation at touchdown exceeds the tip-over threshold, i.e.\ $\abs{\phi^-} > \phi_{\rm tol}$, or \emph{translational failure} when the forward velocity at take-off drops below a chosen threshold, e.g.\ $v_{\rm G,x}^+<0.01$.}
\label{fig:problemsetup}
\end{figure}

We model the runner in the sagittal plane as a rigid body (\fig~\ref{fig:problemsetup}a) of mass $m$, radius of gyration $r_g$, i.e.\ moment of inertia $\I = m r_{g}^2$ about its center of mass, and radius $r_\ell$ (leg length). All quantities are in units such that $m = 1,r_\ell = 1$ and the acceleration due to gravity $g=1$.
See section~\ref{sec:notation} for notation used in this paper.

\subsection{Aerial and stance phases}
\label{sec:running dynamics}
A single step is comprised of an aerial and a stance phase.
The aerial phase is modeled as a drag-free projectile in uniform gravity.
Stance involves two successive parts: a passive collision with the ground followed by an active push-off.
The passive collision is two-dimensional and parameterized by two coefficients of restitution $\en$ along the normal to the ground and $\et$ along the tangent to the ground.
The active push-off applies a linear impulse $\vec{J}_{\rm imp}$ at the contact point P and a rotational impulse $J_{\phi}$ at the center of mass G.
The governing dynamical equations are,
\begin{subequations} \label{eqn:dynamics of running}
\begin{eqnarray}
\label{eqn:passive}
\text{passive collision:}\ &  
&\vec{v}_{\rm P}^c = \begin{pmatrix}
\epsilon_{\rm t} & 0 \\
0 & -\epsilon_{\rm n} \end{pmatrix} \vec{v}_{\rm P}^-, \\
\label{eqn:ang-mom}
& &{H}_{\rm /P}^c - {H}_{\rm P}^- = 0,\\
\label{eqn:pushoff}
\text{push-off:}\ & 
&\vec{v}_{\rm G}^+ = \vec{v}_{\rm G}^c + \vec{J}_{\rm imp}, \quad \I\omega^+ = \I\omega^c + J_{\rm imp,t} + J_{\phi}, \\
\label{eqn:flight}
\text{flight:}\ & 
&\ddot{x}_{\rm G}(t) = 0,\quad \ddot{y}_{\rm G}(t) = -1,\quad \ddot{\phi}(t) = 0,\ \text{and}\\
\label{eqn:flight initial conditions}
\text{initial conditions:}\ & 
&\begin{pmatrix}x_{\rm G}(0) \\ y_{\rm G}(0)\\ \phi(0) \end{pmatrix} = \begin{pmatrix} x_{\rm G}^+ \\ y_{\rm G}^+\\ \phi^+ \end{pmatrix},\; \begin{pmatrix} \dot{x}_{\rm G}(0) \\ \dot{y}_{\rm G}(0)\\ \dot{\phi}(0) \end{pmatrix} = \begin{pmatrix} v_{\rm G,x}^+ \\ v_{\rm G,y}^+\\ \omega^+ \end{pmatrix}.
\end{eqnarray}
\end{subequations}
Horizontal and vertical positions are denoted by $x$ and $y$, respectively, orientation by $\phi$, velocity by $\vec{v}$, angular velocity by $\omega$, moment of inertia by $\I$,
and angular momentum by ${H}$.
Superscript `$-$' denotes variables immediately preceding the collision, `$c$' after the passive collision and `$+$' after the active push-off.
Subscripts P and G refer to quantities associated with the foot and center of mass, respectively.

The mechanical state of the runner is parameterized by the center of mass positions $(x_{\rm G}, y_{\rm G})$, body orientation $\phi$, and their respective velocities $(v_{\rm G,x}, v_{\rm G,y})$ and $\omega$.
Because the stance is assumed to be instantaneous, the velocities may change discontinuously but the position and orientation remain constant during stance.
The instantaneous stance assumption also implies that unmodeled finite forces such as gravity or air-drag do not contribute to the impulse on the runner. 
However, the active rotational impulse ${J}_{\phi}$ applied at the center of mass G captures the effects of varying posture over stance and the changing center of pressure on the ground.
We examine this approximation and its implications in the discussion.


\subsection{Stance: passive collision}
\label{sec:eten}
The runner may control the passive collisional impulse with the ground by varying the parameters $\en$ and $\et$.
Because the collisional impulse passes through P, the angular momentum of the runner about the contact point ${H}_{\rm /P}$ does not change (equation~\eqref{eqn:ang-mom}) and governs the change in the angular velocity $\omega$ due to the collision.

The passive normal collision can vary from perfectly inelastic to perfectly elastic, and is parameterized by the normal coefficient of restitution $0\le\en\le 1$.
The fraction $\epsilon_{\rm n}^2$ models elastic energy stored and recovered during stance.
When $\en = 0$, the runner is completely dissipative and $\en = 1$ implies perfectly energy conserving. 

The tangential coefficient of restitution $\et$, parameterizes the tangential impulse when the foot undergoes a collision with the ground.
Modulation of $\et$ is the feature that distinguishes open-loop versus anticipatory strategies in our model.
Consider the example where the runner modulates $\et$ by varying the tangential foot speed at touch-down.
The open-loop runner would vary the foot speed by assuming that the terrain is flat and that its own mechanical state matches that of a perfectly periodic and steady speed runner.
On a rough terrain the mechanical state and the terrain slope vary from step-to-step.
Thus the intended tangential collision $\etr$ and the actual tangential collision $\et$ may not be equal for the open-loop runner.
The anticipatory runner would use information of the terrain's slope in the oncoming step and its own mechanical state to make sure that the intended and actual foot speed at touchdown match.
Thus the actual and the intended (controlled) tangential collision $\etr$ are equal for the anticipatory runner.
The relationship between $\etr$ and $\et$ are therefore,
\begin{equation}
\et = 
\begin{cases}
\etr \left(\frac{v_{\rm x0}}{v_{\rm P,t}^-}\right) &\text{: open-loop}, \\
\etr &\text{: anticipatory},\\
\end{cases}
\label{eqn:omega-etr}
\end{equation}
where $v_{\rm P,t}^-$ is the tangential velocity of P just before landing and $v_{\rm x0}$ is the steady forward velocity of the center of mass on flat ground.
The two policies are identical on flat ground and when the body has no angular velocity prior to landing.
A numerical examination of the relationship between $\etr$ and $\et$ on rough terrain, is presented in \supp~\ref{ext-sec:et-vs-etc}.

\subsection{Stance: active push-off}
\label{sec:pushoff}

Stance ends with the application of an active, linear push-off impulse $\vec{J}_{\rm imp}$ at the contact point P and an active angular push-off impulse $J_\phi$ at the center of mass G.
We constrain these impulses so that in the absence of external perturbations or other disturbances the runner is perfectly periodic and remains upright ($\phi(t)=0$) on flat ground.
Importantly, once the impulses are chosen for flat ground, they are not allowed to vary step-to-step on any other terrain to reflect the absence of active feedback control.
Together, these conditions imply that that the active push-off impulses $\vec{J}_{\rm imp}$ and $J_\phi$ depend only on $\en$, $\et$, $v_{\rm x0}$ and $v_{\rm y0}$, and no other parameters, according to
\begin{subequations}
\label{eqn:jimp}
\begin{eqnarray}
\label{eqn:jimp1}\vec{j}_{\rm imp} &=& \begin{pmatrix} v_{\rm x0} \\ v_{\rm y0} \end{pmatrix} - \begin{pmatrix} (\et + \frac{1-\et}{1 + I_{\rm /G}})v_{\rm x0} \\ -\en v_{\rm y0} \end{pmatrix},\\
\label{eqn:jimp2}\quad j_\phi &=& 0.
\end{eqnarray}
\end{subequations}
Thus the center of mass of a periodic runner on flat ground has a constant forward speed $v_{\rm x0}$ and vertical speed $v_{\rm y0}$ at every step.

On rough terrains, there are two options for defining the application of the invariant linear impulse on every step.
First, the impulse vector $\vec{J}_{\rm imp}$ may be held constant in every step with respect to gravity (x-y frame in \fig~\ref{fig:problemsetup}a), which we call the \emph{lab-fixed} push-off policy.
Second, the impulse vector may be held constant in every step with respect to the normal direction to the terrain at the point of contact (t-n frame in \fig~\ref{fig:problemsetup}b), which we call a \emph{terrain-fixed} push-off policy.
The terrain-fixed policy may be considered a better approximation of what animals do, because the normal to the terrain and the leg orientation are often coupled, whereas leg orientation at contact and gravity may vary from step to step.
Implicit in preferring the terrain-fixed policy is the assumption that joint torques to apply forces are planned in an ego-centric (body-fixed) frame of reference.
For the disc-like model of a runner that we use, the terrain-fixed and body-fixed policies are identical.
Detailed expressions for the velocities in the stance phase, as well as expressions for $\vec{J}_{\rm imp}$ under both push-off policies are given in \supp~\ref{ext-sec:modeldetails}.
We present an complete analysis of the lab-fixed push-off policy in \supp~\ref{ext-sec:labpushoff} and focus on the terrain-fixed policy in the main paper.

\section{Monte Carlo simulations}
\label{sec:mc}
A sagittal plane runner can only fail by two modes, when the body orientation exceeds a chosen threshold (\emph{orientational failure}), or by failing to move forward any longer (\emph{translational failure}).
We choose the orientational threshold $\phi_{\rm tol}$ as the angle of tilt to passively topple a human who is standing with their feet apart in a pose resembling double-stance in walking.

We perform \mc simulations on step-like and undulating rough terrains to estimate the statistics of failure for both open-loop and anticipatory runners.
Stability is quantified by the mean steps to failure, like previous studies of rough terrain walking \citep{Byl2009}.

The terrain is modeled as a piecewise linear interpolation of an underlying random grid.
The grid points are separated by a distance $\lambda$ and the heights $h$ of the grid points are chosen from a uniform random distribution (table~\ref{table:parameters}, \sec~\ref{sec:terrain model}).
A linear interpolation between the grid points yields a terrain with random variations in both slope and height.
Corners at grid point implies an indeterminate slope, and we therefore define an effective slope at the grid points by interpolating the slope before and after the point (details in \sec~\ref{sec:landing}).
Parameter values that represent a human-like runner (table~\ref{table:parameters}) are used for all \mc simulations, unless indicated otherwise.

\begin{table}
\centering
\begin{tabular}{*9c}
  \toprule
  \multicolumn{5}{c}{Runner} & \multicolumn{2}{c}{Terrain} & \multicolumn{2}{c}{Monte Carlo} \\
  \addlinespace
  $I_{\rm /G}$ & $\epsilon_{\rm n}$ & $\phi_{\rm tol}$ & $v_{\rm x0}$ & $v_{\rm y0}$ & $h$ & $\lambda$ & $M$ & MAX \\
  \cmidrule(lr){1-5}\cmidrule(lr){6-7}\cmidrule(lr){8-9}
  0.17 & 0.63 & $\pi/6$ & 0.96 & 0.26 & $\thicksim\mathcal{U}(-0.03, 0.03)$ & 0.1 & $10^5$ & $10^3$ \\
  \bottomrule
\end{tabular}
\caption{Parameter values for a human-like runner. 
Units are chosen such that $m=1$, $g = 1$, and $r_\ell = 1$.
Parameters describing the runner are discussed in \sec~\ref{sec:model}.
The heights $h$ at grid points defining the terrain are chosen from a uniform distribution over the range $[-0.03,0.03]$ (\sec~\ref{sec:terrain model}).
The ensemble size used in the \mc simulations is $M$, and MAX is the number of steps to which the runner is simulated.
All runners failed before reaching MAX.
We elaborate on the choice of these values in \sec~\ref{sec:simulation}. 
}\label{table:parameters}
\end{table}

\subsection{Open-loop runners on rough terrains}
\label{sec:openloop}

Open-loop runners always fail through an orientational instability on rough terrains regardless of the energy dissipated per step (\fig~\ref{fig:Large perturbations}a,d). 
The open-loop runners with human-like inertia and size take $9.6 \pm 4.1$ steps (mean $\pm$ standard deviation) before tumbling while only 1\% of the runners fall within 3 steps (\fig~\ref{fig:Large perturbations}a). 
Decreasing $\en$ from 1 to 0 increases the mean steps to failure by just 2 steps  (\fig~\ref{fig:Large perturbations}d).
Thus, dissipating more energy per step in the normal collision has minimal influence on stability. 
The tangential collision, parameterized by $\etr$, has little or no influence because the contour lines of the mean steps to failure are nearly parallel to the $\etr$ axis (\fig~\ref{fig:Large perturbations}d).
Therefore, energy dissipation or modulating the tangential collision are both ineffective stabilization strategies for purely open-loop runners.

\subsection{Open-loop runners on step-like terrains}
\label{sec:step-like terrains}
The purely open-loop runner remains stable on step-like terrains that are piecewise flat and possess only height variations (\fig~\ref{fig:Large perturbations}b).
This is because forward and vertical dynamics are decoupled on piecewise flat terrains, and hence the open-loop runner does not fall as long as the step height is smaller than the apex height of the aerial phase.
This result suggests a foot placement strategy for running on any rough terrain, namely to aim to land on flat patches of the ground so that stability is maintained with little reliance on feedback control.
However, such a strategy would require visual surveying of the terrain up ahead and planning the location of foot falls.
Increased footfall probability on flat regions of the terrain, along with low probability of footfalls on highly sloped regions of the terrain may be evidence for such a foot placement strategy in experiments.

\subsection{Effect of terrain geometry}
\label{sec:height distribution}
\begin{figure}
\centering
\includegraphics{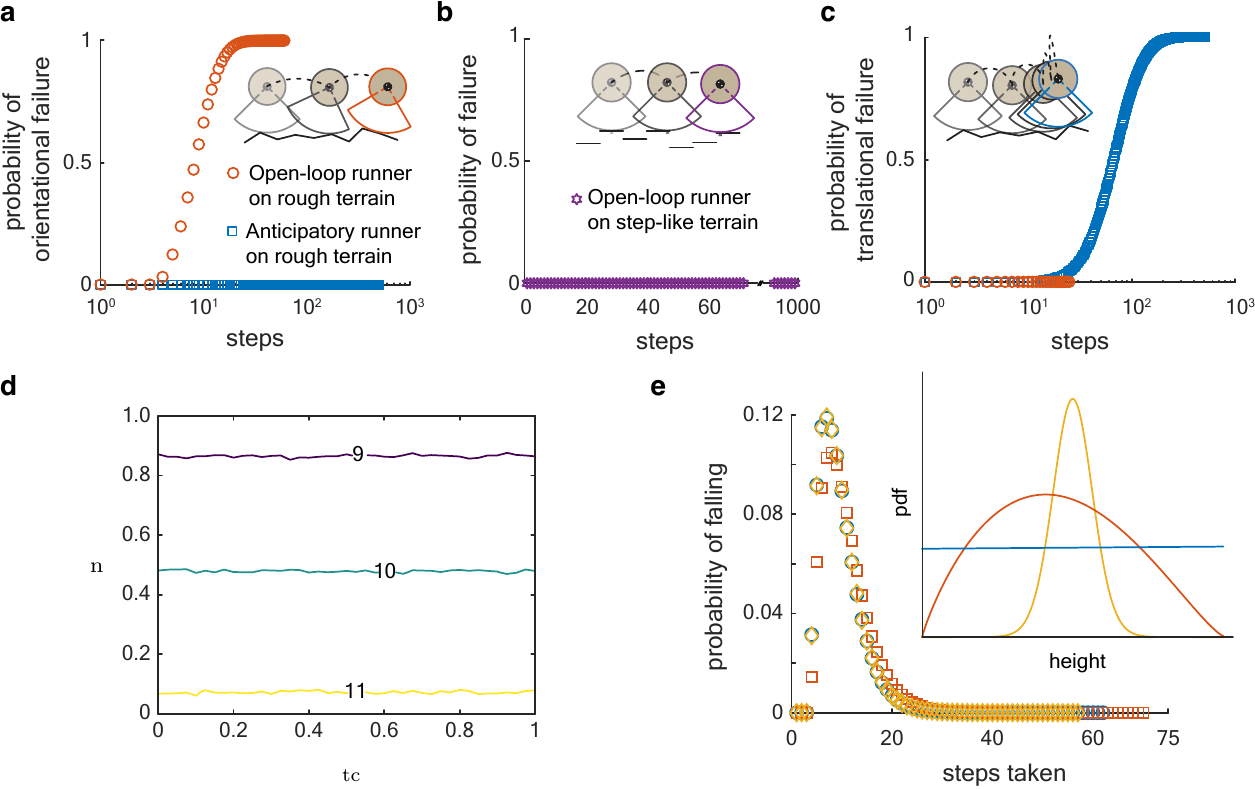} 
\caption{The effect of the tangential collision, energy dissipation and terrain geometry on running stability for a human-like runner, found using \mc simulations.
{\bfseries a}, Open-loop runners with $\etr = 0$ (orange circles) lose orientational stability on the rough terrain while anticipatory runners with $\etr = 1$ (blue squares) maintain orientation.
{\bfseries b}, On the step-like terrain, open-loop runners (purple star) maintain forward speed and orientation as the probability of failure, orientational or translational, is zero.
Open-loop and anticipatory runners are identical on step-like terrains
{\bfseries c}, Anticipatory runners slow down on the rough terrain, eventually completely losing forward speed.
Whereas human-like open-loop runners also lose forward speed, they lose orientational stability before completely losing forward momentum.
{\bfseries d}, A contour plot of mean steps taken by open-loop runners as a function of $\etr$ and $\en$ finds that contours are approximately parallel to the $\etr$ axis, while the steps taken increases with decreasing $\en$.
{\bfseries e}, Steps to failure distributions for human-like open-loop runners on rough terrain with height distributions for the grid points drawn from von~Mises (yellow diamond, mean = 0, $\kappa = 6$), Beta (orange square, $\alpha = 1.9,\, \beta = 2.3$) and uniform distributions (blue circle).
The inset shows the probability density functions for the three distributions used to generate the terrain: von~Mises (yellow), Beta (orange) and uniform (blue).
The distributions were scaled and shifted such that mean height $=0$, and range = $0.060r_\ell$ (table 1, \sec~\ref{sec:simulation}).}
\label{fig:Large perturbations}
\end{figure}

The exact step-to-step variation in the terrain's height and slope depend on the distribution function used to generate the random terrain (\fig~\ref{fig:Large perturbations}e, inset).
However, we find that the distribution underlying the rough terrain has little effect on the distribution of the steps to failure (\fig~\ref{fig:Large perturbations}e) when assessed using three different functions to generate rough terrains: von~Mises, uniform and beta.
Runners took $9.6\pm4.1$ (mean $\pm$ std.\ dev.) steps before tumbling on terrains characterized by von~Mises and uniform distributions, and $10.8\pm4.7$ steps on the terrains characterized by the beta distribution.
All steps to failure distributions are unimodal, but skewed.
A Markov model for the step-to-step dynamics (\supp~\ref{ext-sec:levy}) lends insight into the nearly invariant shape of the steps-to-failure distribution.
The insensitivity may arise from the terrain roughness being uncorrelated from step-to-step (terrain's correlation length $\lambda \ll 1$), and thus the net effect of the perturbations resembles a Gaussian noise process that is propagated by the dynamics of running.

\subsection{Anticipatory runners on rough terrains: tangential collisions}
\label{sec:anticipate}

\begin{figure}
\centering
\includegraphics{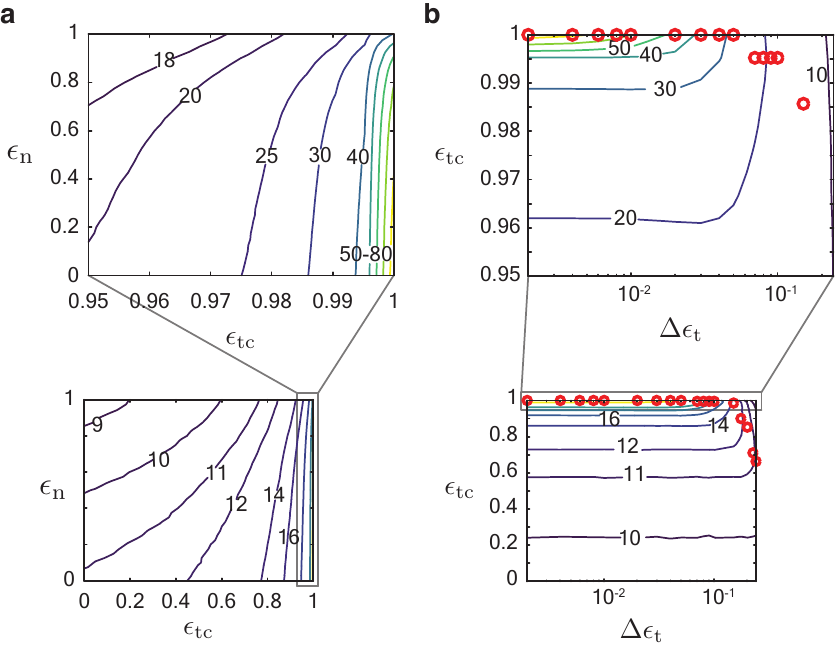}
\caption{Effect of tangential collisions and energy dissipation on running stability for anticipatory runners.
In each panel, the contour plot of mean steps taken over the entire range of independent parameters is shown together with a zoom-in of contours that are bunched together.
{\bfseries a}, The contour plot of mean steps taken by anticipatory runners as a function of $\etr$ and $\en$ shows that contours are bunched close together around $\etr \simeq 1$, with the maximum steps taken at $\en = 0, \etr = 1$ (top plot) and minimum at $\en = 1, \etr = 0$ (bottom plot).
{\bfseries b}, Effect of noise in $\etr$. Contour plot of mean steps taken as a function of $\etr$ and $\Delta\epsilon_{\rm t}$. 
The optimal $\etr$ (red circles) is shown for each value of $\Delta\epsilon_{\rm t}$ simulated.
}
\label{fig:anticipatory}
\end{figure}

Runners that use anticipatory strategies to control the tangential passive collision maintain orientational stability if they entirely avoid tangential collisions using $\etr=1$.
But these runners eventually fail by completely losing forward momentum (\fig~\ref{fig:Large perturbations}c).
Recall that because of the active push-off, the loss of forward momentum is not simply a break of symmetry by the passive tangential collision.
Through a more careful analysis, we find that the mean slope encountered by the runners is positive and not zero, i.e.\ the terrain preferentially impedes the forward momentum (\supp~\fig~\ref{ext-fig:terrainsmooth}d).
For human-like parameters, anticipatory runners take $75 \pm 40.9$ (mean $\pm$ std.\ dev) steps before completely losing forward momentum and only 1\% of the runners stop moving forward within 15 steps.
In contrast, over 80\% of the open-loop runners fail within 15 steps.

For the anticipatory runner, permitting tangential collisions $\etr < 1$ induces orientational failures and the mean steps to failure decreases.
For example, with $\en=0$ the mean steps to failure when $\etr=1$ is 85 and decreases to 20 when $\etr=0.95$ (\fig~\ref{fig:anticipatory}a).
A 5\% decrease in $\etr$ caused an over three-fold decrease in the mean steps to failure.
Importantly, the dominant mode of failure switches from translational failures to orientational failures (\supp~\fig~\ref{ext-fig:ant-runners}a).
At $\etr = 0$ and independent of $\en$, the anticipatory and open-loop strategies are identical.
Thus the anticipatory runner substantially improves stability by avoiding tangential collisions.

Increasing energy dissipation in the normal collision increases the number of steps taken by the anticipatory runner.
For example, at $\etr = 1$, where runners only fail by losing forward speed, increasing energy dissipation by changing from $\epsilon_{\rm n} = 1$ to $\epsilon_{\rm n} = 0$, increases the mean number of steps taken before failure by two-fold, from 40 to 85 (\fig~\ref{fig:anticipatory}a). 
Away from $\etr = 1$, energy dissipation has a smaller effect on stability.
When $\etr \approx  0$, the anticipatory runners resemble the open-loop runners and the mean steps to failure increases by only 2 steps despite $\en$ decreasing from 1 to 0 (\fig~\ref{fig:anticipatory}a).
Thus, for the anticipatory runner using $\etr \approx 1$, increasing energy dissipation in the direction normal to the terrain is an effective means to improve stability, unlike for the open-loop runner.

\subsection{Noise in anticipatory strategies}
\label{sec:noisy retract}
The sensitivity of the steps to failure with respect to tangential collisions prompts an examination of the effect of stochasticity in how a runner may control the tangential collision.
After all, no runner can exactly control the tangential collision from step-to-step.
For example, errors in sensing the terrain profile as well as motor noise may prevent accurate implementation of a desired $\etr$.
We model such sources of noise in controlling the tangential collision as
\begin{subequations}
\begin{eqnarray}
\label{eqn:etnoise}
\epsilon_{\rm t, noisy} &=& \etr + \Delta\et \eta, \\
\text{where }\eta &\thicksim& \mathcal{U}[-1,1], \, \Delta\et \in \mathbb{R}.
\end{eqnarray}
\end{subequations}
The uniformly distributed zero-mean random variable $\eta$ models random step-to-step noise in $\etr$ and $\Delta\epsilon_{\rm t}$ is the noise intensity.

We find that incurring tangential collisions ($\etr<1$) is optimal when there is non-zero noise ($\Delta\epsilon_{\rm t}>0$).
This is unlike the noiseless anticipatory runner whose optimum is $\etr = 1$.
However, noise in controlling tangential collisions does affect stability and the mean steps to failure are severely reduced (\fig~\ref{fig:anticipatory}b).
For example, compared to a noiseless human-like runner, the mean steps to failure drops nine-fold for a runner with noise intensity $\Delta\epsilon_{\rm t} = 0.1$, and the optimum $\etr$ decreases by 1\%  to $\etr = 0.99$ (\fig~\ref{fig:anticipatory}b).
Additional noise in the tangential collision of open-loop runners reduces the number of steps taken, but does not alter the dependence of steps taken on $\etr$ (\supp~\ref{ext-sec:open-loop noise}).
Therefore, for anticipatory runners, noise in controlling the tangential collision implies that incurring a slight tangential collision is optimal but at the cost of stability.

\subsection{Predictions for $\epsilon_{\rm t}$ in experiments}
\label{sec:eth}
A main finding of our analyses is the importance of minimizing tangential collisions with the ground when running on rough terrains.
But measuring $\et$ on rough terrains is challenging because it needs a well-defined point of contact under the foot, precise knowledge of the terrain's slope in 3D at that point, and measurement of the reaction force along that tangent.
To facilitate comparisons with experimental data, we consider an easier to measure correlate of $\et$ via the parameter $\hat{\epsilon}_{\rm t}$ that is defined as
\begin{equation}
\hat{\epsilon}_{\rm t} = 1 - \frac{\Delta v_x}{v_x},
\end{equation}
where $\Delta v_x/v_x$ is the fraction of the forward momentum of the runner lost due to the passive collision.
On perfectly flat terrain, $\et=\hat{\epsilon}_{\rm t}$.

\begin{figure}
\centering
\includegraphics{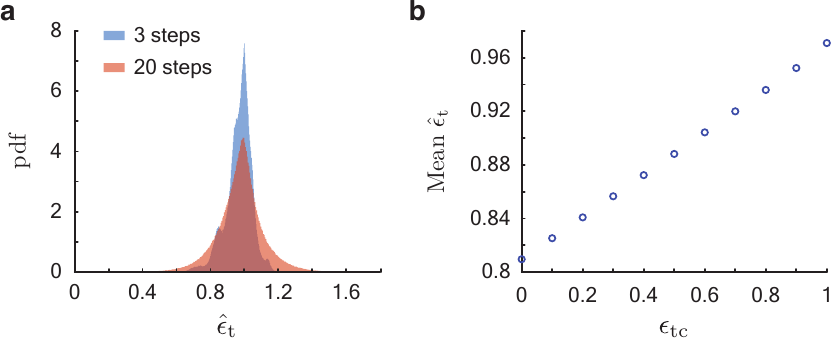}
\caption{Estimated tangential coefficient of restitution $\hat{\epsilon}_{\rm t}$ for anticipatory runners using \mc simulations with an ensemble size of $10^6$. {\bfseries a}, Probability density function of $\hat{\epsilon}_{\rm t}$ for human-like anticipatory runners with $\etr = 1$ on rough terrain after 3 steps and after 20 steps. While the standard deviation almost doubles between the two distributions shown here (\supp~\fig~\ref{ext-fig:stepwise-eth}c), the mean of the distribution converges by 3 steps (\supp~\fig~\ref{ext-fig:stepwise-eth}b). {\bfseries b}, Mean $\hat{\epsilon}_{\rm t}$, converges by 3 steps for all values of $\etr$ (\supp~\fig~\ref{ext-fig:stepwise-eth}b), and is always less than 1, ranging from 0.81 at $\etr = 0$ to 0.97 at $\etr = 1$.}
\label{fig:ethat}
\end{figure}

In the \mc simulations, $\hat{\epsilon}_{\rm t}$ is characterized by a distribution that evolves with increasing steps (\fig~\ref{fig:ethat}a, \supp~\fig~\ref{ext-fig:stepwise-eth}a).
The dependence of $\hat{\epsilon}_{\rm t}$ on steps taken arises because the runner is slowing down, and thus $v_x$ and consequently $\Delta v_x$ change from step-to-step.
But, the mean of $\hat{\epsilon}_{\rm t}$ appears to converge to a constant after just 3 steps for all values of $\etr$ (\supp~\fig~\ref{ext-fig:stepwise-eth}b).
Importantly, mean $\etht$ increases linearly with $\etr$ (\fig~\ref{fig:ethat}b) and is this a reliable correlate of the true tangential collision.
However, $\etht$ has a reduced range; mean $\etht = 0.81$ at $\etr = 0$, and mean $\etht = 0.97$ at $\etr = 1$.
The standard deviation of the distributions converges to a value between 0.05 and 0.1 by approximately 10 steps for most values of $\etr$ except when $\etr \to 1$ (\fig~\ref{fig:ethat}a, \supp~\fig~\ref{ext-fig:stepwise-eth}b).
For comparison, reported values of $\hat{\epsilon}_{\rm t} $ from experiments with human runners on flat and two rough terrains are $0.94\pm0.01$ (mean $\pm$ standard deviation) identically \citep{Dhawale2018ASB}. 
These experimental data are consistent with the prediction that optimal anticipatory runners should maintain $\etr=1$.

\subsection{Modulation of $\et$}

The tangential collisional impulse depends on the speed of the foot at collision and also on how that collisional impulse is transmitted to the center of mass.
For example, the foot collision may not affect the center of mass very much if the intervening joints between the foot and the body are compliant.
The transmission of collisions is treated in terms of sprung and unsprung masses in models of automobiles and in running biomechanics \citep{McGeer1990R}.
If collisional impulses at the foot are faithfully transmitted to the center of mass, the retraction rate $\omega_{\rm ret}$ is related to $\et$ as,
\begin{subequations}\label{eqn:et-omega_ret-reln}
\begin{eqnarray}
\et &=& \frac{\omega_{\rm ret}}{v_{\rm p,t}^-}, \\
\text{where}\ \omega_{\rm ret} &=& \begin{cases} \etr v_{\rm p,t}^-\ \text{: anticipatory},\\
\etr v_{\rm x0} \ \text{: open-loop} \end{cases}.
\end{eqnarray}
\end{subequations}
Using leg retraction to control $\et$ implies that the optimal retraction rate zeros the tangential foot speed at landing.
Equivalently, the foot may be allowed to collide with the ground and yet achieve $\et\approx 1$ by maintaining low stiffness in the leg's joints.


\section{Linear stability analysis}
\label{sec:stability}
For periodic dynamic systems linear stability is defined as the response to small perturbations in the neighborhood of a periodic orbit \citep{Full2002,Holmes2006,Bruijn2013} and analyzed using Floquet theory \citep{Guckenheimer1983,Holmes2006}.
Floquet analysis for the stability of a periodic orbit defines a transverse cross-section to the orbit and a discrete return map from initial conditions on the cross-section back to the same cross-section after a complete period.
The eigenvalues of the return map, called Floquet multipliers, are independent of the chosen cross-section and govern the stability of the periodic solution to small perturbations \citep{Guckenheimer1983}.
Here we consider the anticipatory runner and discuss the open-loop runner in \supp~\ref{ext-sec:jordan} because the unstable modes of both variants are the same. 

The mechanical state of the runner is represented by $\vec{\zeta} = (x, y, \phi, v_x, v_y, \omega)\tr$, where $(x, y)$ and $\phi$ denote the center of mass position and orientation, and $(v_x, v_y)$ and $\omega$ are the respective velocities, all measured in a Newtonian reference frame that translates forward at a constant speed $v_{\rm x0}$.
A steady runner is periodic in this translating Newtonian frame of reference.
We define a transverse cross-section (Poincar\'{e} section) at the apex of the aerial phase ($v_y = 0$) following the approach of \citet{Full2002} and \citet{Seyfarth2003}.
The equations~\eqref{eqn:dynamics of running} yield the step-to-step return map $\vec{f}_{\rm an}$ and its linearization $\vec{\rm T}_{\rm an} $ in terms of a the mechanical state $\vec{\psi}$ in a translating frame according to
\begin{subequations} \label{eqn:floquet analysis}
\begin{eqnarray}
\vec{\psi} &=& (x, y, \phi, v_x, \omega)\tr, \label{eqn:poincare state}\\
\vec{\psi}_{n+1} &=& \vec{f}_{\rm an}\left(\vec{\psi}_n\right), \label{eqn:return map}\\
\Delta\vec{\psi}_{n+1} &=& \vec{\rm T}_{\rm an}\Delta\vec{\psi}_n, \label{eqn:linearized return map}\\
\text{where}\ \Delta\vec{\psi} &=& \vec{\psi}-\vec{\psi}^\ast,\ \vec{\rm T}_{\rm an} = \frac{\partial \vec{f}_{\rm an}}{\partial \vec{\psi}}\bigg|_{\vec{\psi}^*}.\label{eqn:linearized return map definition}
\end{eqnarray}	
\end{subequations}
The Poincar\'{e} map given by equation~\eqref{eqn:return map} has a fixed point at $\vec{\psi}^*=\vec{0}$ when the terrain is flat and corresponds to an exactly periodic runner on flat ground.

\begin{figure}
\centering
\includegraphics{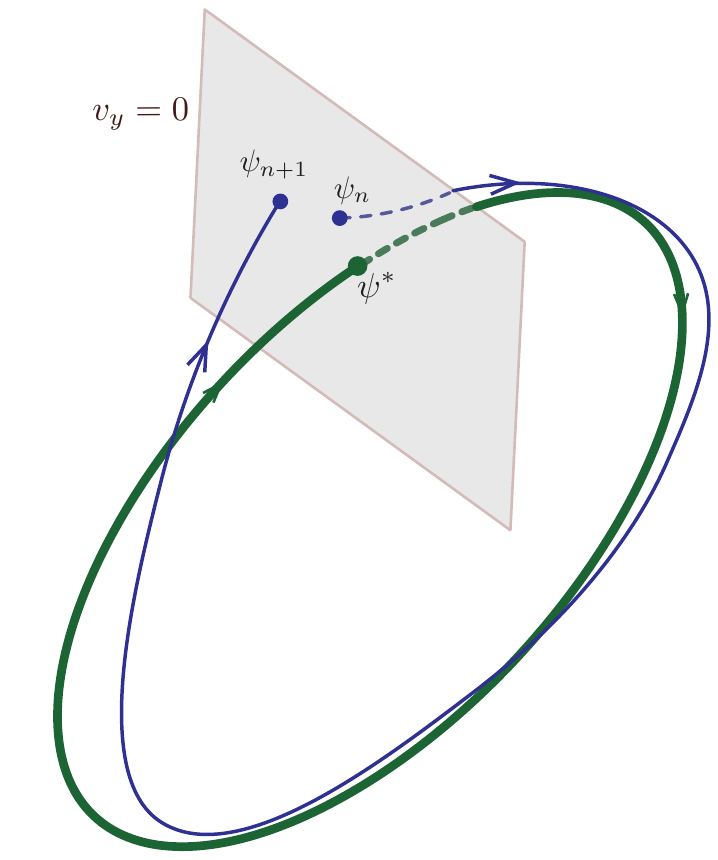}
\caption{Illustration of the trajectory of the runner in state space in a reference frame that is translating along with the runner with velocity $v_{\rm x0}$. 
The runner appears periodic in this reference frame and the runner's mechanical state follows a periodic orbit.
The return map $f_\bullet$ ($\bullet$ is `ol' or `an' for open-loop or anticipatory, respectively) is defined from the apex of the aerial phase ($v_y = 0$) to apex of the following aerial phase.
$\vec{\psi^*}$ is the fixed point of the return map and $\vec{\psi}_n$ is a small perturbation away from the fixed point $\vec{\psi}^*$ at step $n$.
In the next step, $\vec{\psi}_n$ maps to $\vec{\psi}_{n+1}$ at the apex of the following aerial phase under action of the return map $\vec{f}_\bullet$.}
\label{fig:returnmap}
\end{figure}

The linearized return map $\vec{\rm T}_{\rm an}$ has three eigenvalues equal to one and the others are all less than one.
The eigenvalues with magnitude less than one correspond to stable modes so that perturbations along their respective eigenvectors will always decay.
The remaining three eigenvalue are all $\lambda = 1$ with algebraic multiplicity equal to 3 and geometric multiplicity equal to 2.
This implies that there are only two independent eigenvectors corresponding to the three unity eigenvalues and the matrix $\vec{T}_{\rm an}$ is therefore non-diagonalizable.
For non-diagonalizable systems, the Jordan decomposition is used to analyze stability in terms of generalized eigenvectors (\supp~\ref{ext-sec:jordan}), and implies that the modes (eigenvectors) associated with these eigenvalues cannot be decoupled and analyzed independently.

The two eigenvectors $\vec{\nu}_1$, $\vec{\nu}_2$ and the third generalized eigenvector $\vec{\nu}_3$ corresponding to the repeat eigenvalue $\lambda = 1$ span a subspace in which the dynamics of the return map don't simply decay back to the origin.
For a diagonalizable system, any perturbation within this subspace would neither decay nor grow.
However, the non-diagonalizable nature of $\vec{T}_{\rm an}$ leads to the outcome that a perturbation $\Delta\vec{\psi}_0$ within this subspace grows with increasing steps.
The eigenvectors $\vec{\nu}_1$, $\vec{\nu}_2$, $\vec{\nu}_3$, the initial perturbation $\Delta\vec{\psi}_0 $, and its growth after $n$ steps to $\Delta\vec{\psi}_n$ are given by,
\begin{subequations}\label{eqn:generalized instability}
\begin{eqnarray}
	\vec{\nu}_1 &=& \begin{pmatrix} 0 & 0 & 1 & 0 & 0 \end{pmatrix}\tr,\ 
	\vec{\nu}_2 = \begin{pmatrix} 1 & 0 & 0 & 0 & 0 \end{pmatrix}\tr,\ 
	\vec{\nu}_3 = \begin{pmatrix} 0 & 0 & 0 & \frac{-1}{\sqrt{2}} & \frac{1}{\sqrt{2}} \end{pmatrix}\tr,\label{eqn:eigenvectors}\\
	\Delta\vec{\psi}_0 &=& \sum_{k=1}^3 \alpha_k \nu_k,\ \text{and}\\ 
	\Delta\vec{\psi}_n &=& n \alpha_3 \sqrt{2} \en v_{\rm y0} (\vec{\nu}_1 - \vec{\nu}_2) + \Delta\vec{\psi}_0,\ \text{respectively}\label{eqn:subspace perturbation}.
\end{eqnarray}
\end{subequations}
As $n$ grows larger, the asymptotic approximation (denoted by $\approx$) is given by
\begin{equation}
	\Delta\vec{\psi}_n \approx n\, \alpha_3 \sqrt{2}\en v_{\rm y0} \begin{pmatrix} -1 \\ 0 \\ 1 \\ 0 \\ 0 \end{pmatrix}\ \text{where}\ n\gg 1. \label{eqn:lingrowth}
\end{equation}
Only a perturbation of magnitude $\alpha_3$ along $\vec{\nu}_3$ affects stability and leads to a nearly linear growth within the subspace spanned by the eigenvectors $\vec{\nu}_1, \vec{\nu}_2$.
Perturbations along $\vec{\nu}_1$ or $\vec{\nu}_2$ neither grow nor decay  because these represent invariance with respect to rotations and translations of the reference frame, respectively.
A perturbation along the generalized eigenvector $\vec{\nu}_{3}$ may be geometrically viewed as one that conserves the velocity of the contact point on flat terrain but changes the angular momentum of the runner about its center of mass.
Therefore, any perturbation to the angular momentum will affect both orientation and forward speed.

For the special case of the anticipatory runner that completely avoids tangential collisions, the linearized return map $\vec{T}_{\rm an}$ with $\etr = 1$ has eigenvalue $\lambda = 1$ of algebraic multiplicity 4 and geometric multiplicity 2, and one eigenvalue with $|\lambda| < 1$.
The eigenvectors $\vec{\nu}_1, \vec{\nu}_2$ and the generalized eigenvectors $\vec{\nu}_3,\vec{\nu}_4$ associated with $\lambda = 1$ form a basis for a subspace within which an initial perturbation $\vec{\psi}_0$ grows linearly with the number of steps $n$ in a subspace spanned by eigenvectors $\vec{\nu}_1,\vec{\nu}_2$, i.e.\
\begin{subequations}
\begin{eqnarray}
	\vec{\nu}_1 &=& \begin{pmatrix} 0 \\ 0 \\ 1 \\ 0 \\ 0 \end{pmatrix},\ 
	\vec{\nu}_2 = \begin{pmatrix} 1 \\ 0 \\ 0 \\ 0 \\ 0 \end{pmatrix},\ 
	\vec{\nu}_3 = \begin{pmatrix} 0 \\ 0 \\ 0 \\ 0 \\ 1 \end{pmatrix},
	\vec{\nu}_4 = \begin{pmatrix} 0 \\ 0 \\ 0 \\ 1 \\ 0 \end{pmatrix}, \label{eqn:eigenvectors etc=1}\\
	\Delta\vec{\psi}_0 &=& \sum_{k=1}^4 \alpha_k \nu_k,\ 
	\Delta\vec{\psi}_n = n\, (\alpha_3  a_1 \vec{\nu_1} + \alpha_4 a_2 \vec{\nu_2}) + \Delta\vec{\psi}_0,\\
	\Delta\vec{\psi}_n &\approx& n\ 2\en v_{\rm y0} \begin{pmatrix} \alpha_4 \\ 0 \\ \alpha_3 \\ 0 \\ 0 \end{pmatrix} \text{for}\ n \gg 1.
	\label{eqn:etc=1 lingrowth}
\end{eqnarray}
\end{subequations}
A perturbation to angular velocity $\omega$ causes a linear growth in orientation $\phi$, and a perturbation to the linear velocity $v_x$ causes a linear growth in position $x$.
However, an anticipatory runner with $\etr = 1$ avoids angular velocity perturbations due to the terrain altogether, i.e.\ $\alpha_3 = 0$.
Therefore, only forward speed is affected due to the remaining unstable mode $\vec{\nu}_4$.

Although there are no unstable eigenvalues with magnitude greater than one, we find that the dynamics of running lead to an unstable growth with increasing steps.
The growth due to non-diagonalizability of the return map is linearly proportional to the number of steps, and not geometric as is the case for simple unstable eigenvalues.
Importantly, the primary effect of the instability is to affect the forward speed and orientation, consistent with the numerical simulations that use finite perturbations and nonlinear dynamics.
Also in agreement with the simulations, the only instability is translational when $\etr=1$.

\section{Scaling analysis of the orientational failure mode} 
\label{sec:paramdepend}
The mean steps to failure depends on many parameters, but none of the parameters separately predict the failure statistics (\supp~\fig~\ref{ext-fig:general}).
As most runners undergo orientational failures, we investigated whether the amount of body rotation accumulated over a single step due to a terrain slope perturbation would predict failure statistics.

If a runner with a periodic trajectory on flat ground encounters a sloped terrain of angle $\theta$, the orientation $\phi_\bullet$ at the next landing will no longer be vertical. 
This orientation $\phi_\bullet$ accumulated over one step depends on the take-off vertical velocity $v_{y,\bullet}^+$ via the aerial phase time $2v_{y,\bullet}^+$, and take-off angular velocity $\omega_\bullet^+$, as $\phi_\bullet = 2v_{y,\bullet}^+\omega_\bullet^+$.
The subscript `$\bullet$' is a placeholder for `ol' or `an' as the orientation change depends on whether the runner is purely open-loop (ol) or employs anticipatory (an) control.
We hypothesize that the mean steps to failure $N_\bullet$ is a function of the orientational threshold $\phi_{\rm tol}$ and the orientation change over a single step $\phi_\bullet$ alone, i.e.\ $N_\bullet = s_\bullet(\phi_{\rm tol},\phi_\bullet)$.
Substituting the form of $s_\bullet(\phi_{\rm tol},\phi_\bullet)$ derived in \supp~\ref{ext-sec:levy}, we show that the mean steps to failure $N_\bullet$ is predicted to scale according to,
\begin{equation}
\label{eqn:paramdep}
N_\bullet \sim \frac{\phi_{\rm tol}}{\phi_\bullet},
\end{equation}
where the expression for $\phi_\bullet$ is given in \supp~\eq~\eqref{ext-eqn:phi-bullet-compl}.

\begin{figure}[!thb]
\centering
\includegraphics{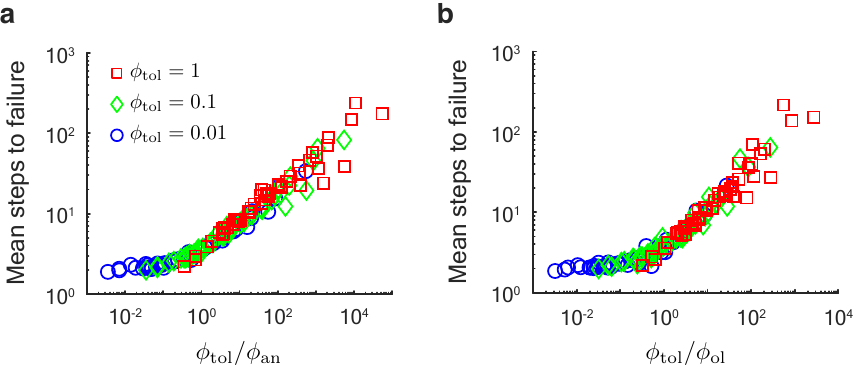}
\caption{Generalizing results from section~\ref{sec:openloop} and section~\ref{sec:anticipate} to a wider range of physical and terrain parameters. 
Mean steps to failure from the Monte Carlo simulations is plotted against {\bfseries a}, $\phi_{\rm tol}/\phi_{\rm an}$ and {\bfseries b}, $\phi_{\rm tol}/\phi_{\rm ol}$ for different values of $\phi_{\rm tol}$.
The mean steps to failure depend mostly on a single dimensionless parameter $\phi_{\rm tol}/\phi_{\bullet}$.
All simulation parameters were varied independently in these simulations.
But, for clarity, only variations in $\phi_{\rm tol}$ are identified with different marker types.}
\label{fig:Smallperturbation}
\end{figure}

The mean steps to failure in simulations performed with many different parameter values (\supp~\ref{ext-sec:bodyrot}) are well-approximated by a single function of a dimensionless parameter $\phi_{\rm tol}/\phi_\bullet$ (\fig~\ref{fig:Smallperturbation}).
The collapse of the simulation data highlights that the spin accumulated in one step due to a single perturbation (\eq~\eqref{eqn:paramdep}) captures the fundamental principle underlying orientational failures.
Importantly, this dimensionless parameter collapses the simulation data better than any individual parameter (\supp~\fig~\ref{ext-fig:general}).
Thus, the single parameter $\phi_{\rm tol}/\phi_\bullet$ quantifies stability of runners of different sizes and mass distributions.

\begin{figure}[!thb]
\centering
\includegraphics{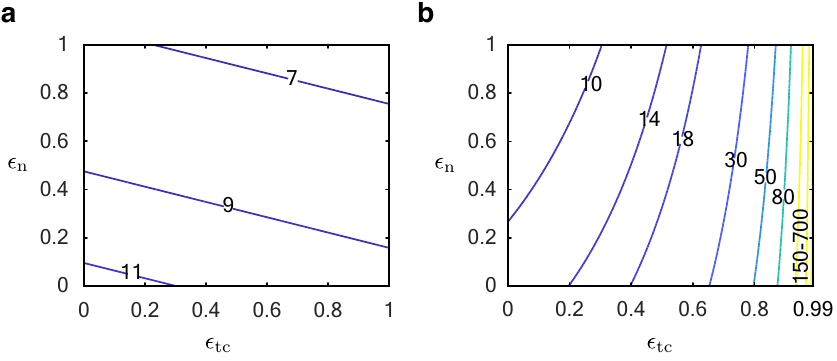}
\caption{Contour plots of {\bfseries a}, $\phi_{\rm tol}/\phi_{\rm ol}$ and {\bfseries b}, $\phi_{\rm tol}/\phi_{\rm an}$ as a function of $\en$ and $\etr$ reveal that a single parameter captures the dependence of mean steps to failure of both open-loop runners (\fig~\ref{fig:Large perturbations}d) and anticipatory runners (\fig~\ref{fig:anticipatory}a) on the collision parameters $\en$ and $\etr$.
Recall that the only controllable parameters for the runners in these simulations are $\en$ and $\etr$.
The complete expression for $\phi_\bullet$, shown in \supp~\eq~\eqref{ext-eqn:phi-bullet-compl}, was used to generate these plots with parameter values drawn from table 1, and $\phi_{\rm tol} = 1$.
For the anticipatory runner, we restricted the maximum value of $\etr$ to 0.99. For higher values of $\etr$ orientational failures are rare and thus not accounted for by $\phi_{\rm tol}/\phi_\bullet$.
}
\label{fig:asymptotic analysis contours}
\end{figure}

The dimensionless parameter $\phi_{\rm tol}/\phi_\bullet$ also captures the parametric dependence of mean steps to failure on $\etr$ and $\en$ as seen from comparing the contour plots of $\phi_{\rm tol}/\phi_\bullet$ shown in \fig~\ref{fig:asymptotic analysis contours} against that of the direct simulations in \fig~\ref{fig:Large perturbations}d and \fig~\ref{fig:anticipatory}a.
The dependence of mean steps to failure on $\etr$ and $\en$ for small slopes of the terrain is understood using a series expansion of $\phi_\bullet$ in terms of $\theta$ as given by,
\begin{subequations}\label{eqn:series exp}
\begin{eqnarray}
\label{eqn:phi-ol}
\phi_{\rm ol} &=& \bigg(\frac{2v_{\rm y0}^2}{1+\I}\bigg)\theta + \bigg(\frac{3-\I}{(1+\I)^2}+\frac{4\I}{(1+\I)^2}\etr + \frac{2}{1+\I}\en \bigg) v_{\rm x0}v_{\rm y0}\, \theta^2 + O(\theta^3),\\
\label{eqn:phi-an}
\nonumber \phi_{\rm an} &=& \bigg(\frac{2v_{\rm y0}^2}{1+\I}(1-\etr)\bigg)\theta + \bigg(\frac{3-\I}{(1+\I)^2} + \frac{5\I-3}{(1+\I)^2}\etr -\frac{4 \I}{(1+\I)^2}\etr^2 + \\
& & \frac{2 (1-\etr)}{1+\I}\en \bigg)v_{\rm x0}v_{\rm y0}\,\theta^2 + O(\theta^3).
\end{eqnarray}
\end{subequations}
For the open-loop strategy, neither of the collision parameters, $\en$ or $\etr$, appear in the linear (leading order) term.
When using an anticipatory strategy, the tangential collision parameter $\etr$ appears to leading order.
The normal collision parameter $\en$ affects the second order dependence on $\theta$ for both strategies.
These show why it is impossible to avoid orientational failures for the open-loop strategy, but may be avoided when using the anticipatory strategy by choosing $\etr=1$ and $\en=0$.

For the open-loop runner, $\phi_{\rm tol}/\phi_{\rm ol}$ is independent of $\en$ and $\etr$ to first order in $\theta$ (\eq~\eqref{eqn:phi-ol}).
Hence the contours in \fig~\ref{fig:asymptotic analysis contours}a (which resemble the contours in \fig~\ref{fig:Large perturbations}d from the \mc simulations) show a weak dependence on $\en$ and $\etr$ that arises from the $\theta^2$ term in \eq~\eqref{eqn:phi-ol}.
The parameter $\phi_{\rm ol}$ is smallest when $\en = \etr = 0$, and largest when $\en = \etr = 1$.
For a human-like runner, $\I\ll 1$ (table 1), and thus the $\theta^2$ term in \eq~\eqref{eqn:phi-ol} can be reduced to $(3+2\en)v_{\rm x0}v_{\rm y0}\theta^2$, with no dependence on $\etr$ at the asymptotic limit of $\I\ll 1$.
The asymptotic analysis of $\phi_{\rm ol}$ therefore explains why the contours of mean steps to failure in the \mc simulations are nearly parallel to the $\etr$ axis and increase only slightly when $\en$ is decreased (\fig~\ref{fig:Large perturbations}d).

For the anticipatory runner, the first order term in the expansion depends on $\etr$ (\eq~\ref{eqn:phi-an}), unlike the case for the open-loop runner (\eq~\eqref{eqn:phi-ol}).
Nearly perfect anticipation corresponds to $\etr \to 1$.
At this limit $\phi_{\rm an} \to 0$ and thus $N=\phi_{\rm tol}/\phi_{\rm an} \to \infty$, explaining why the contours of mean steps to failure in the \mc simulations are tightly bunched together in the neighborhood of $\etr = 1$ (\fig~\ref{fig:anticipatory}a) and nearly parallel to the $\en$ axis.
Like for the open-loop runner, $\phi_{\rm an}$ also shows a dependence on $\en$ only in the $\theta^2$ term of the power series expansion in \eq~\ref{eqn:phi-an}.
As $\en$ decreases so does $\phi_{\rm an}$, and thus $\phi_{\rm tol}/\phi_{\rm an}$ increases, capturing the trend observed in the \mc simulations where decreasing $\en$ increases steps taken for the anticipatory runner (\fig~\ref{fig:anticipatory}a).
For the anticipatory runner, unlike the open-loop runner, the $\en$ dependence is coupled to $\etr$, and thus the sensitivity of the parameter $\phi_{\rm tol}/\phi_\bullet$ to changes in $\en$ depend on the value of $\etr$.
The limit of $\etr = 0$, where $\phi_{\rm an} = \phi_{\rm ol}$ has already been discussed above, for the open-loop runner.
To analyze the case where $\etr \to 1$, we approximate $\phi_{\rm an}$ in the limit where $\I << 1$ (e.g.\ human-like runners) as
\begin{equation}
\phi_{\rm an} \approx 2v_{\rm y0}^2(1-\etr)\theta + (1-\etr)(2\en+3)v_{\rm x0}v_{\rm y0}\theta^2.
\label{eqn:phi-an-approx}
\end{equation}

To understand the dependence of the mean steps to failure $N$ on $\en$ and $\etr$, we consider the limit of small angles of the terrain slope $\theta \ll 1$.
Using \eq~\eqref{eqn:phi-an-approx}, and for small $\theta$ we find that mean steps to failure $N = \phi_{\rm tol}/\phi_{\rm an}$ and its sensitivity to changes in $\en$ are given by
\begin{subequations}
\begin{eqnarray}
\label{eqn:N approx}
N &=& \left(\frac{1}{1-\etr}\right)\frac{\phi_{\rm tol}}{v_{\rm y0}\theta}(2v_{\rm y0} - 3v_{\rm x0}\theta - 2v_{\rm x0}\theta\en),\\
\label{eqn:N approx sensitivity en}
\frac{\partial N}{\partial \en} &=& -\left(\frac{1}{1-\etr}\right) \frac{2 \phi_{\rm tol} v_{\rm x0}}{v_{\rm y0}}.
\end{eqnarray}
\end{subequations}
Therefore, $N$ is more sensitive to changes in $\en$ when $\etr \to 1$.
This resembles \fig~\ref{fig:anticipatory}a where the mean steps to failure from the \mc simulations increases significantly as $\en$ is reduced when $\etr \to 1$, as opposed to when $\etr \to 0$ where there is much lesser sensitivity of the mean steps to failure with respect to changes in $\en$.

Improving running stability by increasing mean steps to failure helps provide more time for feedback driven corrections in real-world runners.
The analysis of mean-steps to failure in the simplified runners without any feedback ability suggests that increasing $\phi_{\rm tol}$ and decreasing $\phi_\bullet$ are both effective strategies to negotiate rough terrains.
Therefore, besides altering $\etr$ and $\en$ in order to increase $\phi_{\rm tol}/\phi_\bullet$ as already discussed, increasing $\I$ and reducing $v_{y0}$ also improves stability.

\section{Discussion}
\label{sec:discussion}
We show that purely open-loop strategies with no feedback control cannot stabilize sagittal-plane dynamics during running.
Such open-loop runners fail primarily by losing orientational stability and tumbling.
Using an anticipatory strategy to eliminate tangential collisions with the ground eliminates orientational instabilities but leads to a steady slowing down of the runner.
However, on a step-like piecewise flat terrain the open strategy is sufficient to stabilize the runner without losing forward speed, and so is the anticipatory strategy.
If an anticipatory strategy is implemented noisily, i.e.\ the tangential collisions are low but not entirely eliminated, the runner suffers orientational instabilities.
However, both the orientational and the translational instabilities are weak when using an anticipatory strategy and the growth of the instability is only linearly proportional to the number of steps taken and not a higher power.
The exact number of steps to failure depend on many parameters including the inertial, geometry, collision parameters and the thresholds in orientation and speed for failure.
These large set of parameters may be combined into a single dimensionless parameter that captures the failure statistics, when can also guide the morphological design of stable runners.

\subsection*{Impulsive stance assumption}
An impulsive stance phase implies that the stance impulse is defined, but not the detailed time history of forces.
Thus the model may be used study the dynamics and stability over multiple steps, but it cannot be used to find the actuation patterns that would achieve the desired impulse.
Such simplified models may used to specify the desired collisional and push-off impulses as constraints that should be be met.
More detailed models could then be used to calculate the stance force profiles as a constrained search or optimization problem.


The model also ignores the impulse due to the finite forces of gravity, because stance is treated as instantaneous.
Relaxing the assumption of instantaneous stance implies that body-weight affects the body's angular momentum about the contact point according to,
\begin{equation}
\label{eqn:ang mom finite stance}
{H}_{\rm /P}^+-{H}_{\rm /P}^- = \int\limits_{0}^{T_{\rm stance}} {M}_{\rm /P}(t)\, dt,
\end{equation}
where $T_{\rm stance}$ is the stance duration and ${M}_{\rm /P}(t)$ is the time-varying moment of the body weight about the contact point P.
The torque due to gravitational forces about the contact point is proportional to body weight and the time-varying horizontal distance from the center of mass to the contact point.
The gravitational contribution is zero for a symmetric stance, and highest for the most asymmetric stance.
Assuming a constant forward speed during stance and a 20$^\circ$ touchdown angle, the maximum change in angular momentum about the contact point, i.e.\ the integral in \eq~\eqref{eqn:ang mom finite stance}, is $\lvert\Delta{H}_{\rm /P}\rvert \lessapprox 0.15$ in the same dimensionless units as before.
For a typical human runner, the resultant orientation change in a single step is $\Delta\phi \lessapprox 0.01$, negligibly small compared to the influence of the terrain.
Thus, ignoring torques induced by gravity has minimal impact on our conclusions.

\begin{figure}[!thb]
\centering
\includegraphics{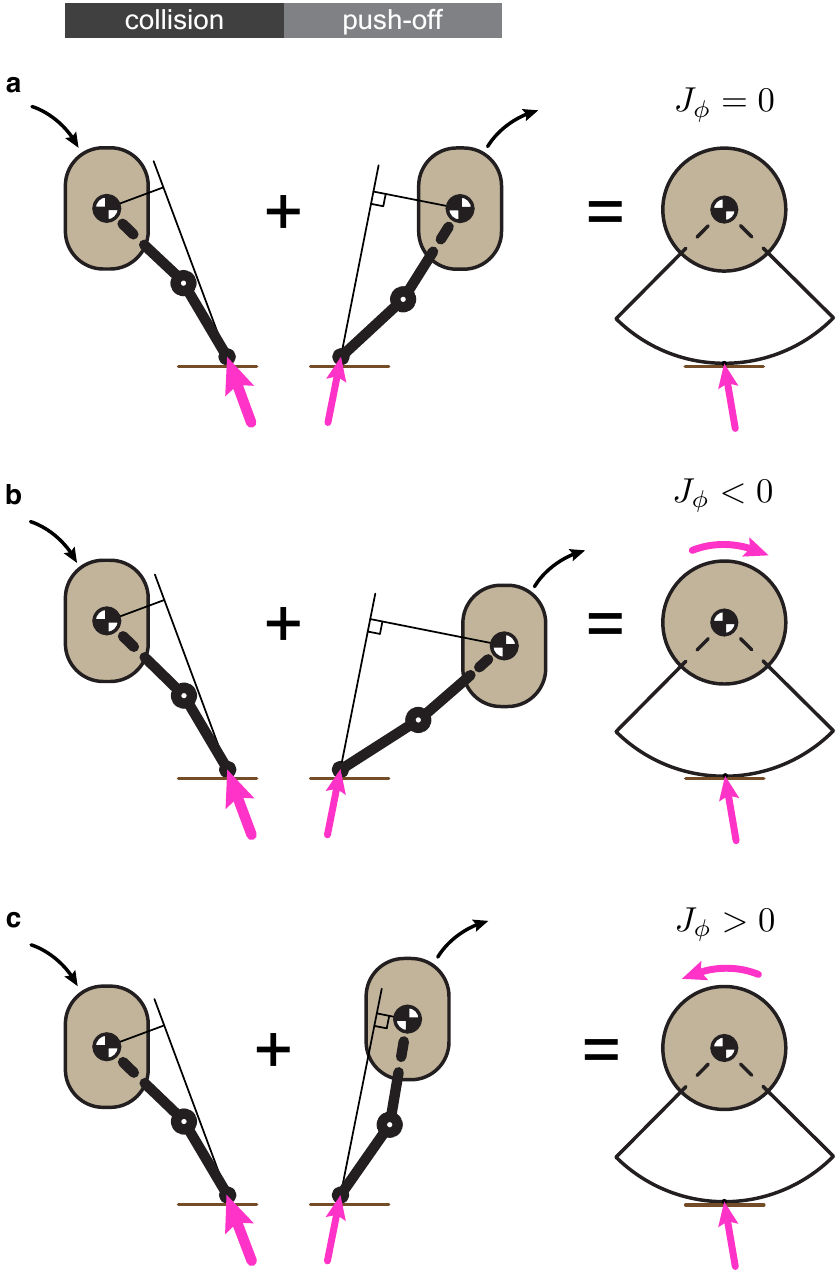}
\caption{Equivalence between a runner using a freedom finite stance with internal degrees of freedom versus an infinitesimal stance with an applied torque impulse.
The contact of the general runner is represented as an impulse due to the collision at touchdown (left, pink vector) plus a net impulse due to the push-off (right, pink vector).
In this example, by varying duration of stance, the runner can selectively vary the lever arm of the push-off impulse and thus the angular impulse about its center of mass without altering the linear impulse.
The net effect of a finite stance and change in configuration is therefore captured by a linear impulse at the point of contact (pink vector at ground) and an angular impulse at the center of mass ($J_{\phi}$).
}
\label{fig:finite infinitesimal equivalence}
\end{figure}

A finite stance duration and the associated change in configuration allows a runner to control the body's sagittal plane angular momentum, independently from the forward and upward linear momenta.
This may be understood in terms of breaking the symmetry of the stance phase or applying a large forward impulse and yet having the net ground reaction impulse pass through the center of mass (no contribution to angular momentum).
To not lose such control when considering an impulse stance, we permit the application of an arbitrary angular impulse $J_{\phi}$ at the center of mass during push-off (equation~\eqref{eqn:pushoff}).
Thus having a finite stance duration and change of body configuration over stance is equivalent to $J_{\rm phi}$ (\fig~\ref{fig:finite infinitesimal equivalence}).
The angular impulse also provides a means to accommodate torques due to a finite base of support and a moving center of pressure during stance.
However, recall that the constraint that the applied active push-off impulses should lead to perfectly periodic gait on flat terrain implies that $J_{\phi}=0$.
In our model, both open-loop and anticipatory runners slow down on rough terrain.
Regaining forward speed needs a feedback controller, and then the additional control authority offered by $J_{\phi}$ would be necessary to vary forward speed without affecting the body's angular momentum or vertical momentum.

\subsection*{Point contact assumption}
Another limitation arises from considering a point-like contact that cannot capture effects associated with the spatial extent of the foot.
These effects include the spatial filtering of terrain roughness and the application of a net torque about the initial contact point.
The inclusion of $J_{\phi}$ in the model captures the application of torques, but there is no explicit means of incorporating the ability of the foot to act as a spatial filter \citep{Sharbafi2017}.
Therefore, careful consideration should be given to the spatial frequency (wave number) of the roughness of the terrain when using a model with a point contact.

\subsection*{Timescale for feedback corrections}

Open-loop runners with human-like parameters have a 99\% chance of taking at least 3 steps without failing by exceeding the orientation threshold, while anticipatory runners ($\etr = 1$) can take upto 15 steps with the same probability of completely losing forward momentum. 
This implies that the open-loop runner employing the slowest sensory modality (visual feedback delay $\approx200$~ms \citep{vanBeers2002}) has 7 feedback cycles to correct for instabilities at endurance running speeds of 3m/s (step period $\approx500$~ms \citep{Cavagna1964}), with only an approximately 1\% chance of an orientational failure.
Thus, while sensory feedback is required to run on rough terrains, timescales associated with sensory feedback delays do not limit the runner's ability to maintain stability because of the nature of the instability.
Furthermore, employing an appropriate anticipatory strategy ($\etr = 1$) eliminates the orientational instability entirely, thereby further extending the timescale over which feedback is necessary.

\subsection*{Leg retraction}

Analyses of running models with leg mass suggest that optimal retraction rate is defined by stability demands, although these studies were limited to step-like terrains \citep{Karssen2015}.
Experiments with runners on flat ground which measure the angle of the foot's velocity vector with respect to the ground suggest that foot velocity is perhaps not modulated in the manner we hypothesize \citep{Blum2010}.
In the study by \citet{Blum2010}, the mean angle made by the subjects' foot velocity vector with the ground was 165$^\circ$, whereas our prediction based on zero tangential speed of the foot at touchdown would imply that the angle should be 90$^\circ$.
Given these differences, we propose that the low values of $\hat{\epsilon}_{\rm t}$ for human runners observed by \citet{Dhawale2018ASB} may result from joint stiffness modulation in the leg rather than precise control of the foot speed through leg retraction.
Modulating foot and leg stiffness allows the runner to minimize the tangential collision and yet employ leg retraction strategies that accomplish other goals such as hypothesized by \citet{Seyfarth2003} and \citet{Birn2014}.

\subsection*{Energy dissipation}

Besides leg retraction, energy dissipation may also aid in stability based on studies of walking \citep{Kuo1999,Donelan2001} and running \citep{Daley2006,Arellano2011,Arellano2012}.
Our model shows that while increasing energy dissipation in the direction normal to the terrain does increase the number of steps taken for open-loop and anticipatory runners, dissipating energy in the tangential collision is detrimental to stability.
However, whether energy dissipation helps or hinders depends on the details of what is meant by ``open-loop''.
For example if the runner uses a \emph{lab-fixed} push-off policy instead of the \emph{terrain-fixed} push-off described in the main text, dissipating energy in the normal direction is also detrimental to orientational stability (\supp~\ref{ext-sec:labpushoff}).
Thus the hypothesized trade-off between energy consumption and stability is not universally true in our models.
Our results are consistent with experiments on running birds encountering sudden terrain drops, as the birds do not always dissipate energy on the perturbation step \citep{Daley2006}.
Our results might provide a means to understand why the increase in energy consumption for humans running on step-like terrains is only 5\% \citep{Voloshina2015}.
We find that open-loop strategies are sufficient to maintain stability on step-like terrains and additional energy expenditure provides little added benefit.



\subsection*{Implications of scaling analysis to body plan of animals}

The single parameter $\phi_{\rm tol}/\phi_\bullet$ that predicts mean steps to failure (\fig~\ref{fig:Smallperturbation}) generalizes our results beyond runners with human-like parameters and can be used as a criteria to assess a runner's stability.
This is because runners with a larger $\phi_{\rm tol}/\phi_\bullet$ should be able to maintain orientation for a greater number of steps in the absence of sensory feedback control.
We have discussed how energy storage in the direction normal to the terrain ($\en$) and tangential collision modulation ($\etr$) affects $\phi_{\rm tol}/\phi_\bullet$ in \sec~\ref{sec:paramdepend}, and now turn to the implications the parameter has on how body morphology and mass distribution affect stability.

The parameter $\phi_{\rm tol}$ is the maximum angle of tilt the runner can accumulate before it falls.
By employing larger (base/height) ratios, i.e.\ adopting a landscape rather than a portrait orientation when viewed in the sagittal plane, animals can increase $\phi_{\rm tol}$ and thereby increase $\phi_{\rm tol}/\phi_{\rm \bullet}$.
Quadrupeds such as cats and dogs, and other adept runners such as cockroaches possess such an aspect ratio.
Another way to increase $\phi_{\rm tol}$ is by increasing the range of motion of the leg with respect to the body.
Even if the body begins to tilt, the ability to place the foot in front of the runner initiates stance and hence allows the runner to correct for body orientation.
In our simulations, the choice of $\phi_{\rm tol}$ value is based on this consideration of the leg angle for humans.
Ostriches are another example of an animal with a portrait orientation but who are adept runners, perhaps in part due to the large of range of motion of their legs.
Penguins, who are not known to be adept runners, occupy the opposite end of $\phi_{\rm tol}$ scale due to possessing a portrait orientation and low range of motion of their legs compared to other bipeds such as humans, turkeys, and ostriches. 

Because $\phi_{\rm tol}/\phi_\bullet \propto \I/v_{\rm y0}^2$ (\eqs~\eqref{eqn:series exp}), lowering take-off angles for a given forward speed would be beneficial to stability.
However, very low take-off angles increase the risk of tripping on rough terrains.
Altering body mass distribution to increase the radius of gyration $r_g$ relative to leg length $r_\ell$ also reduces $\phi_\bullet$ and thereby increases stability.
This can be achieved by increasing distal masses in appendages like arms and legs.
However, increasing distal masses in the leg increases the metabolic cost of running \citep{Myers1985} via increased energetic cost associated with swinging the leg \citep{Marsh2004,Doke2005} and may also lead to potentially injurious collisions.
Alternatively, light legs with a bulky, extended torso or large head, like in horses and bison, also increases $r_g$.
Lastly, tails in animals like cats and lizards, while used for active stabilization and in complex maneuvers like righting reflexes \citep{Jusufi2011,Libby2012}, are yet another means to increase $r_g$, thereby reducing $\phi_\bullet$.
Anticipatory runners further benefit from setting $\et \approx 1$ like observed in humans \citep{Dhawale2018ASB}, thereby drastically reducing $\phi_\bullet$ as discussed in \sec~\ref{sec:paramdepend}.
Thus, the dimensionless parameter $\phi_{\rm tol}/\phi_\bullet$ is qualitatively consistent with the body morphology of adept and poor animal runners and we propose that it can be used as a design criteria for running robots.
\section{Methods}
\label{sec:methods}

\subsection{Simulation methods}
\label{sec:simulation}

All simulations were performed using custom-written C programs. Parameter values given in table 1 are used for simulations of the human-like runner.
These values are chosen for the purpose of illustration, however our qualitative results are not sensitive to these values, and the scaling analysis in section~\ref{sec:paramdepend} addresses the generalization of these numerical results to runners and terrains with varying parameter values.

\subsection{Parameter values for a human-like runner}

The rationale for chosing the human-like parameter values is as follows. 
The moment of inertia value we use is derived from estimates made by \citep{Erdmann1999}, who find that moment of inertia about the center of mass in the sagittal plane is $\approx 13$~kg.m$^2$ for a 75~kg human.
The value for $\epsilon_{\rm n} = 0.63$ corresponds to $\approx40$\% elastic energy stored over one gait cycle, similar to estimates by \citep{Cavagna1964,Cavagna1977b,Alexander1987}.
The orientation bound $\phi_{\rm tol}$ is equal to $\pi/6$ as it approximately half the angle between the legs during double stance in walking.
Forward speed at take-off $v_{\rm x0} = 0.96$ corresponds to 3~m/s for a leg length of 1~m and vertical speed at take-off $v_{\rm y0} = 0.26$ corresponds to $\approx 0.8$~m/s \citep{Dhawale2018ASB}.
Distributions had nearly converged by an ensemble size of $10^4$, hence we simulate for $10^5$ instances (\supp~\ref{ext-sec:convergence}).

\subsection{Terrain model}
\label{sec:terrain model}

The terrain is modelled as piecewse linear.
This is achieved by first defining a one-dimensional grid with fixed grid spacing $\lambda$.
Interpolating heights between the grid points $k$, located at $x_k$ to intermediate points $(x_t,y_t)$ in the $k$th terrain patch, yields a piecewise linear, continuous terrain profile, where terrain slope $m_k$ is discontinous at the grid points,
\begin{subequations}\label{eqn:terrain-model}
\begin{eqnarray}
\text{patch}\ k:\, y_t &=& m_k x_t + c_k,\\ 
\text{where}\ x_t &\in& [x_k, x_{k+1}], \\
\text{continuity condition:}\ m_k x_{k+1} + c_k &=& m_{k+1} x_{k+1} + c_{k+1},
\end{eqnarray}
\end{subequations}
where $m_k$ and $c_k$ are constants within a patch.
Terrain heights at all grid points are distributed according to $\thicksim\mathcal{U}(-0.03, 0.03)$ (table 1).
Our choice of the uniform distribution $\mathcal{U}$ is to improve simulation speed, even though beta distributions described in \fig~\ref{fig:Large perturbations}e most closely matched artificial terrain used in experiments \citep{Dhawale2015DW,Dhawale2018ASB}.
The range of heights $h \in [-0.03, 0.03]$ and grid spacing $\lambda = 0.1$ was chosen to match the artificially constructed rough terrains \citep{Dhawale2015DW,Dhawale2018ASB}.

Step-like terrains with no slope distributions were simulated by picking a height from the probability distribution prior to landing. 
If the chosen landing height was above the apex height of any portion of the runner, we chose another landing height from the distribution. 
The probability of this resampling occuring is $\sim 10^{-4}$. 

\subsection{Calculating ground contact point}
\label{sec:landing}

The aerial phase ends when the runner collides with the ground.
The landing position is determined by solving for the unknown intersection point $x_t$ of the runner's aerial phase trajectory $(x_G,y_G)$ with the condition for tangential contact between runner and ground,
\begin{subequations}\label{eqn:terrain contact}
\begin{eqnarray}
\label{eqn:ballflight}
\text{parabolic flight:}\ y_{\rm G} &=& b_0 + b_1 x_{\rm G} + b_2 x_{\rm G}^2, \\
\text{touchdown:}\ y_{\rm G} &=& y_t + \frac{1}{\sqrt{1+m_k^2}},\; x_{\rm G} = x_t - \frac{m_k}{\sqrt{1+m_k^2}},
\label{eqn:contact}
\end{eqnarray}
\end{subequations}
where $b_0,b_1,b_2$ are constants that define the aerial phase trajectory.
Equations~\eqref{eqn:terrain-model}-\eqref{eqn:terrain contact} solved simultaneously yield a quadratic equation in $x_t$,
\begin{subequations}\label{eqn:landing}
\begin{eqnarray}
\label{eqn:quadratic}
A x_t^2 + B x_t + C &=& 0, \\
\text{where } A &=& b_2, \\
B &=& b1 - 2 b_2 \frac{m_k}{\sqrt{1+m_k^2}} - m_k, \\
C &=& b_0 - \frac{b_1 m_k - 1}{\sqrt{1+m_k^2}} + b_2 \frac{m_k^2}{1+m_k^2} - c_k.
\end{eqnarray}
\end{subequations}
The larger of the two roots of this quadratic is the true landing point $x_{\rm P}$, if the roots are real and the larger of the two roots is greater than $x_k$.
The other real root is always less than $x_k$ and corresponds to the intersection of the aerial phase trajectory with the terrain patch closer to the take-off point.
On flat terrain, the smaller root is the location of the take-off point.
Having solved for $x_{\rm P}$, the position of the center of mass at landing is determined using \eq~\eqref{eqn:contact}.

However, if the runner lands on a grid point, the position of the center of mass appears to be indeterminate as the grid point $x_k$ is associated with two slopes, $m_k$ and $m_{k+1}$.
In fact, we detect a corner collision if the larger root of \eq~\eqref{eqn:quadratic} is less than $x_k$, or if the roots are complex.
Thus, we now know the position of contact point P, $x_{\rm P} = x_k$, but cannot determine $(x_G,y_G)$ at contact using \eq~\eqref{eqn:contact}, since $x_k$ is associated with slopes $m_k$ and $m_{k+1}$.
We determine a unique slope at the point $x_k$ by accounting for the aerial phase trajectory.
Substituting $x_t = x_k$ in \eq~\eqref{eqn:quadratic}, we write \eqs~\eqref{eqn:landing} as a quartic polynomial in unknown $m_k$.
We numerically find all the roots using the Jenkins-Traub algorithm \citep{Jenkins1970} and pick the real root that corresponds to first contact between the ground and runner, i.e.\ when the parabolic trajectory describing the aerial phase is above the ground.

\subsection{Notation}
\label{sec:notation}

Scalars are denoted by italic symbols (e.g.\ $m$ for mass of the runner, $I$ for the moment of inertia), vectors by bold, italic symbols ($\vec{J}_{\rm imp}$ for push-off impulse, $\vec{v}$ for velocity), points or landmarks in capitalized non-italic symbols (such as center of mass G in \fig~\ref{fig:problemsetup}) and capitalized, bold, non-italic symbols for matrices (such as return map matrix $\vec{\rm T}_{\rm an}$). 
Vectors associated with a point, such as velocity of center of mass G are written as $\vec{v}_{\rm G}$, with the uppercase alphabet in the subscript specifying the point in the plane. 
Angular momentum vectors or moment of inertia variables are subscripted with `/A' representing angular momentum or moment of inertia computed about point A, such as $I_{\rm /G}$ representing moment of inertia about center of mass G. 
The component of velocity $\vec{v}_{\rm A}$ (velocity of point A) in the $\hat{x}$ direction is denoted with a subscript `A,t', e.g.\ tangential velocity of the contact point P is written as $v_{\rm P,t}$. 
The symbols $v_{x0}, v_{y0}$ denote the initial forward and vertical velocities of the runner at take-off. 
Variables just before collision with the terrain are denoted by the superscript `-', after passive collision but before push-off by the superscript `c', and just after push-off by the superscript `+'. For example, angular velocity before collision is $\omega^-$, after passive collision is $\omega^c$ and just after push-off is $\omega^+$. 

\section{Authors' Contribution}

MV conceived of the model. 
ND and MV ran the simulations. 
All authors were involved in the analysis of the model; SM, MV and ND performed the linear stability analysis, MV and ND performed the one-step analysis to capture mean statistics, MV, SM and ND did the analysis of the steps to failure distributions.
ND and MV wrote the manuscript, and all authors edited it.

\section{Acknowledgments and funding}
This work was funded by the Human Frontiers Science Program and the Wellcome/DBT India Alliance. 

\section{Competing interests}

We have no competing interests.

\singlespacing



\section*{Supplementary material}

# Supplementary material for 'How to run on rough terrains'

Nihav Dhawale, Shreyas Mandre and Madhusudhan Venkadesan

# Contents

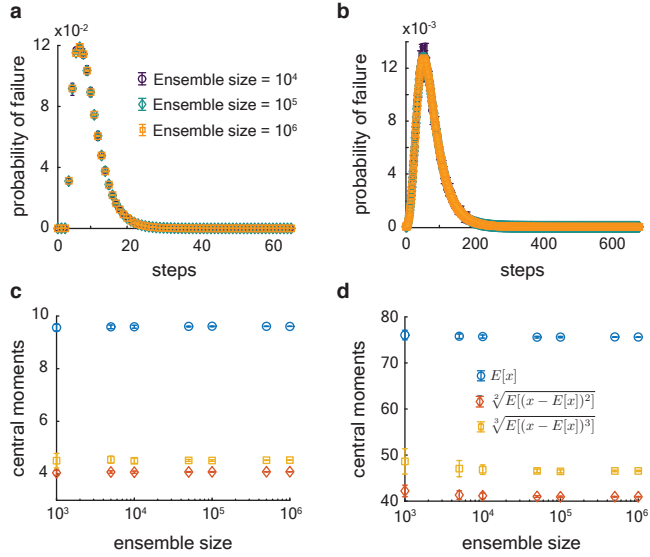

Figure S1: Monte Carlo simulations of open-loop and anticipatory runners with human-like parameter values and varying ensemble sizes. Ten simulations for each ensemble size were performed. Open markers represent the mean value, with error bars showing the standard deviation over the 10 runs. Steps to failure distributions for the **a**, open-loop runner and **b**, anticipatory runner. The y-axis of the plots is the joint probability p(falling at step number i,reaching step number i). The first three central moments as a function of ensemble size for the **c**, open-loop runner and **d**, anticipatory runner are stationary for ensemble sizes greater than $10^4$.

# S1 Convergence of the Monte Carlo simulations

We run Monte Carlo simulations with different ensemble sizes to test for convergence of the steps to failure distributions. Our criteria for convergence is met if the first three moments of the distribution are unchanged (one significant decimal place) as a function of ensemble size. This criteria is met with an ensemble size of $10^4$ (Fig. S1). Hence we perform Monte Carlo simulations in the main text with an ensemble size of at least $10^5$.

# S2 Model Details

## S2.1 Passive collision

A runner with center of mass velocity $\boldsymbol{v}_\text{G}^-$ and angular velocity $\omega^-$, collides with a terrain patch angled at $\theta$ with respect to the horizontal. By simultaneously solving equations describing the collision at the contact point P (main text equation (1a)), and using kinematic constraints associated with a rigid body, the linear velocity $\boldsymbol{v}_\text{G}^c$ and angular velocity $\omega^c$ of the runner after the collision

are computed as,

$$\text{collision law: } \boldsymbol{v}_{\mathrm{P}}^c = \begin{pmatrix} \epsilon_{\mathrm{t}} & 0 \\ 0 & -\epsilon_{\mathrm{n}} \end{pmatrix} \boldsymbol{v}_{\mathrm{P}}^-, \tag{S1a}$$

$$H_{/\mathrm{P}}^c - H_{/\mathrm{P}}^- = 0, \tag{S1b}$$

$$\text{kinematic constraint: } \boldsymbol{v}_{\mathrm{G}} = \boldsymbol{v}_{\mathrm{P}} - \omega \begin{pmatrix} \cos\theta \\ -\sin\theta \end{pmatrix}, \tag{S1c}$$

$$\implies \boldsymbol{v}_{\mathrm{G}}^c = \begin{pmatrix} \frac{(v_{\mathrm{G,x}}^- \cos 2\theta + v_{\mathrm{G,y}}^- \sin 2\theta)A + v_{\mathrm{G,x}}^- B + C\omega^- \cos\theta}{2(I_{/\mathrm{G}}+1)} \\ \frac{(v_{\mathrm{G,x}}^- \sin 2\theta - v_{\mathrm{G,y}}^- \cos 2\theta)A + v_{\mathrm{G,y}}^- B + C\omega^- \sin\theta}{2(I_{/\mathrm{G}}+1)} \end{pmatrix}, \tag{S1d}$$

$$\text{and } \omega^c = \frac{(\epsilon_{\mathrm{t}} + I_{/\mathrm{G}})\omega^- - (1 - \epsilon_{\mathrm{t}})v_{\mathrm{G,t}}^-}{1 + I_{/\mathrm{G}}}, \tag{S1e}$$

where $A = I_{/\mathrm{G}}(\epsilon_{\mathrm{t}} + \epsilon_{\mathrm{n}}) + \epsilon_{\mathrm{n}} + 1$, $B = \epsilon_{\mathrm{t}} I_{/\mathrm{G}} - (I_{/\mathrm{G}} + 1)\epsilon_{\mathrm{n}} + 1$, $C = 2(\epsilon_{\mathrm{t}} - 1)I_{/\mathrm{G}}$, (S1f)

and $v_{\mathrm{G,t}} = v_{\mathrm{G,x}}\cos\theta + v_{\mathrm{G,y}}\sin\theta$. (S1g)

## S2.2 Open-loop push-off strategies

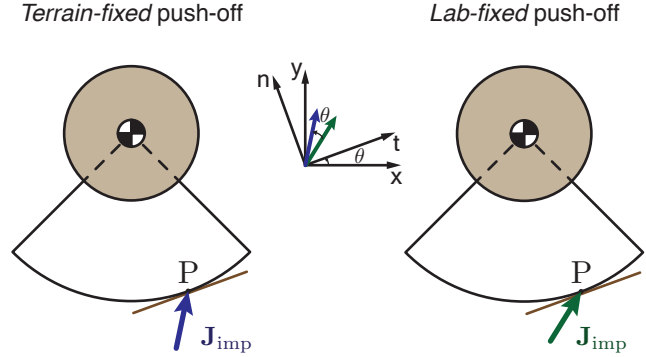

Figure S2: The *terrain-fixed* push-off impulse is the *lab-fixed* push-off impulse rotated by the terrain slope angle $\theta$ about the contact point.

After the passive collision, the push-off impulse $\boldsymbol{J}_{\mathrm{imp}}$ is applied at the contact point P to propel the runner into the flight phase. Recall that $J_\phi = 0$ for the open-loop push policies that we consider (main text equation (1c)). The push-off impulse $\boldsymbol{J}_{\mathrm{imp}}$ leads to discrete changes in the linear velocity of the center of mass $\boldsymbol{v}_{\mathrm{imp}}$, and angular velocity of the runner $\omega_{\mathrm{imp}}$. As discussed in main text section 2.3, a *terrain-fixed* push-off policy or a *lab-fixed* push-off policy both satisfy the periodicity criteria on flat ground, namely $\boldsymbol{v}_{\mathrm{G,steady}}^+ = (v_{x0}, v_{y0})^{\mathrm{T}}$, and angular velocity is $\omega^+ = 0$, but differ in the effect on rough terrain. The linear velocity $\boldsymbol{v}_{\mathrm{G}}^+$ and angular velocity $\omega^+$ at take-off

under these push-off policies are,

$$\boldsymbol{v}_{\text{G}}^+ = \boldsymbol{v}_{\text{G}}^c + \boldsymbol{v}_{\text{imp}}, \quad \omega^+ = \omega^c + \omega_{\text{imp}}, \tag{S2a}$$

$$\text{where } \boldsymbol{v}_{\text{imp}} = \begin{cases} \boldsymbol{R} \begin{pmatrix} \frac{I_{/\text{G}}(1-\epsilon_t)v_{\text{x}0}}{I_{/\text{G}}+1} \\ (1-\epsilon_n)v_{\text{y}0} \end{pmatrix} & : \textit{terrain-fixed}, \\ \begin{pmatrix} \frac{I_{/\text{G}}(1-\epsilon_t)v_{\text{x}0}}{I_{/\text{G}}+1} \\ (1-\epsilon_n)v_{\text{y}0} \end{pmatrix} & : \textit{lab-fixed}, \end{cases} \tag{S2b}$$

$$\omega_{\text{imp}} = \begin{cases} \frac{(1-\epsilon_t)v_{\text{x}0}}{I_{/\text{G}}+1} & : \textit{terrain-fixed}, \\ \frac{(1-\epsilon_t)v_{\text{x}0}\cos\theta}{I_{/\text{G}}+1} + \frac{(1-\epsilon_n)v_{\text{y}0}\sin\theta}{I_{/\text{G}}} & : \textit{lab-fixed}, \end{cases} \tag{S2c}$$

$$\text{and } \boldsymbol{R} = \begin{pmatrix} \cos\theta & -\sin\theta \\ \sin\theta & \cos\theta \end{pmatrix}, \tag{S2d}$$

is the rotation matrix between the *lab-fixed* (x-y) reference frame and the *terrain-fixed* (t-n) reference frame (Fig. S2).

## S3 Open-loop runners on rough terrain

### S3.1 The open-loop passive collision

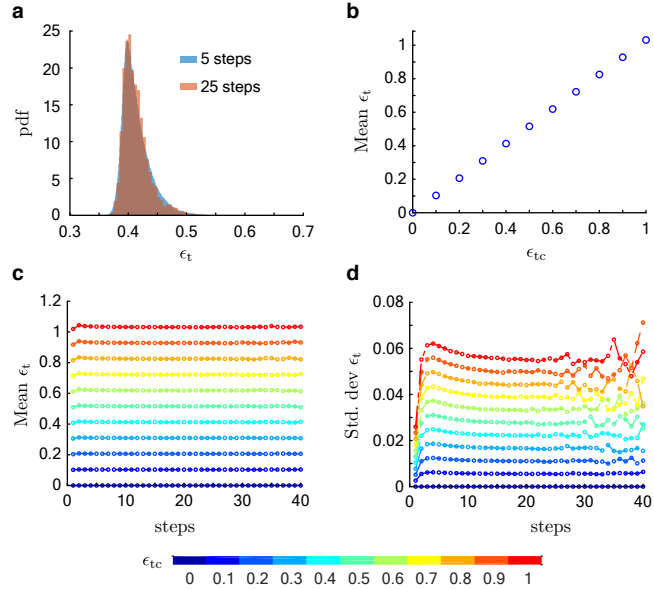

Figure S3: Tangential collisions in open-loop runners with human-like parameters on rough terrain. Monte Carlo simulations performed with $10^6$ runners. **a**, The probability density function of $\epsilon_t$ after 5 steps and 25 steps for $\epsilon_{\text{tc}} = 0.4$. **b**, Steady-state mean $\epsilon_t$ increases linearly with $\epsilon_{\text{tc}}$. **c**, Mean $\epsilon_t$ as a function of step number for different values of $\epsilon_{\text{tc}}$ (shown in the color bar below) converges by approximately 3 steps. **d**, Standard deviation of $\epsilon_t$ as a function of step number for different values of $\epsilon_{\text{tc}}$ also reaches a steady-state. However, the rate of convergence is slower at higher values of $\epsilon_{\text{tc}}$. The fluctuations at higher step number ($\gtrsim 25$) are because we are probing the tails of the steps to failure distributions.

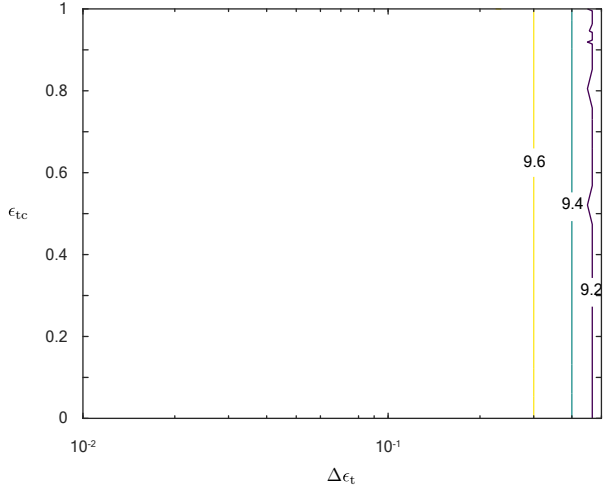

Figure S4: Effect of additive noise in $\epsilon_{tc}$ for open-loop runners with human-like parameters. Contour plot of mean steps to failure as a function of $\epsilon_{tc}$ and $\Delta\epsilon_t$. The mean steps to failure is constant as a function of $\epsilon_{tc}$ and $\Delta\epsilon_t$ below $\Delta\epsilon_t = 0.2$, and shows a small decrease with increasing $\Delta\epsilon_t$ when $\Delta\epsilon_t > 0.2$.

The tangential collision parameter $\epsilon_t$ for open-loop runners varies from step-to-step on rough terrains (main text equation (2)) and is thus characterized by distributions (Fig. S3a) that evolve according to running dynamics described in main text equation (1). These distributions achieve stationarity as seen in the representative case of the $\epsilon_t$ distribution with $\epsilon_{tc} = 0.4$, that appears to converge by 5 steps (Fig. S3a). Mean $\epsilon_t$ converges by 3 steps to values that are just a little larger than the corresponding $\epsilon_{tc}$ values (Fig. S3c), and show a linear dependence on $\epsilon_{tc}$ (Fig. S3b). The shape of the distribution is affected by the $\epsilon_{tc}$ value as the standard deviation of the converged $\epsilon_t$ distribution increases with $\epsilon_{tc}$ (Fig. S3d).

The stability benefits of higher mean $\epsilon_t$ values (Fig. S3b) are counteracted by corresponding higher fluctuations in $\epsilon_t$ (Fig. S3d), leading to larger step-to-step fluctuations in body angular momentum. Hence for open-loop runners, mean steps to failure shows a weak dependence on $\epsilon_{tc}$ value. In main text section 5, we arrive at this same conclusion via an asymptotic analysis of the series expansion of the orientation change over a single step caused by an unexpected terrain slope perturbation.

### S3.2 Additive noise in open-loop strategies

Additive noise in the tangential collision (main text section 3.5) of open-loop runners has little effect on mean steps to failure (Fig. S4) compared to the same noise intensity applied to anticipatory runners (main text Fig. 3b). For open-loop runners with human-like parameters, there is no optimal value of $\epsilon_{tc}$ (Fig. S4) unlike the case with anticipatory runners where slight tangential collisions are optimal (main text Fig. 3b). The relative insensitivity of failure statistics to added noise for the open-loop runner is possibly because $\epsilon_t$ varies significantly on rough terrains even in the absence of added noise (Fig. S3).

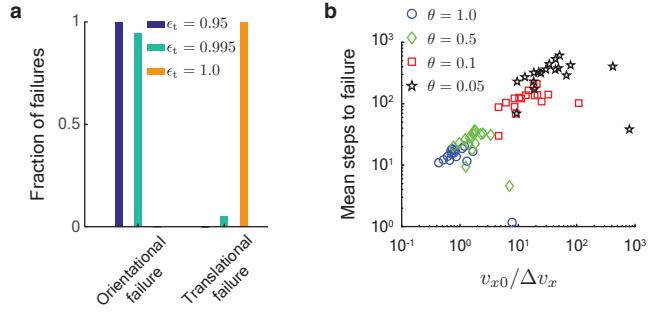

Figure S5: Anticipatory runners on rough terrain. **a**, Relative frequency of failure modes for anticipatory runners with $\epsilon_n = 0$ on rough terrain. The orientation failure bound $\phi_{\text{tol}} = \pi/6$ in these simulations and other parameters are the same as those for the human-like runner (main text table 1) **b**, Mean steps to failure as function of $v_{x0}/\Delta v_x$ for the same range of parameters shown in Fig. S7, except that $\epsilon_{\text{tc}} = 1$ here.

## S4 Anticipatory runners on rough terrain

In main text section 3.4 we describe how zeroing the tangential collision impulse stabilizes orientation, but deviations from this strategy quickly leads to a loss in stability. To underscore how quickly failure modes change, consider that anticipatory runners with $\epsilon_{\text{tc}} = 1$ fail exclusively due to translational instabilities while anticipatory runners with human-like parameters and $\epsilon_{\text{tc}} = 0.95$ fail exclusively due to orientational instabilities (Fig. S5a). Thus, the predominant failure mode switches quickly from translational failures to orientation failures close to $\epsilon_{\text{tc}} = 1$; even runners with $\epsilon_{\text{tc}} = 0.995$ predominantly undergoing orientational failures (Fig. S5a).

The mean steps to failure for an anticipatory runner with $\epsilon_{\text{tc}} = 1$ is captured by a single parameter $v_{x0}/\Delta v_x$ (Fig. S5b). The logic is similar to that in main text section 5 where the scaling analysis is for orientational failures, while here the failure mode is due to losing translational stability (main text Fig. 1c). For a runner landing on sloped ground with forward speed $v_{\text{G,x}}^- = v_{x0}$, the loss in forward speed due to the terrain perturbation is $\Delta v_x = v_{x0} - v_{\text{G,x}}^+$ over a single step. Following the logic outlined in main text section 5, the mean steps to failure $N$ should scale as,

$$N \sim v_{\text{x0}}/\Delta v_x, \tag{S3a}$$
$$\text{where } \Delta v_x = \sin\theta(v_{y0}(\cos\theta - 1 + \epsilon_n(\cos\theta + 1)) + v_{x0}\sin\theta(1 + \epsilon_n)). \tag{S3b}$$
$$\text{Power series: } \Delta v_x = 2\epsilon_n v_{y0}\theta + v_{x0}(1 + \epsilon_n)\theta^2 + O(\theta^3). \tag{S3c}$$

Equation (S3c) is the power series expansion for $\Delta v_x$ about $\theta = 0$ to second order in $\theta$. The power series analysis suggests that shallower flight trajectories (reducing $v_{y0}$) and increasing energy dissipation (reducing $\epsilon_n$) increases the mean steps to translational failure. Results from the Monte Carlo simulations in main text Fig. 3a find that reducing $\epsilon_n$ increases mean steps taken for anticipatory runners with $\epsilon_{\text{tc}} = 1$, consistent with equations (S3b)-(S3c).

### S4.1 Forward collision parameter $\hat{\epsilon}_t$ as a function of $\epsilon_{\text{tc}}$

The forward collision parameter $\hat{\epsilon}_t$ is characterized by a distribution as described in main text section 3.6. The distributions of $\hat{\epsilon}_t$ with $\epsilon_{\text{tc}} = 0.5$ after 3 steps and 20 steps are relatively unchanged (Fig. S6a) in comparison to the distribution of $\hat{\epsilon}_t$ with $\epsilon_{\text{tc}} = 1$ (main text Fig. 4a) where the distribution broadens significantly more. The mean of the distribution for all values of $\epsilon_{\text{tc}}$ converges by

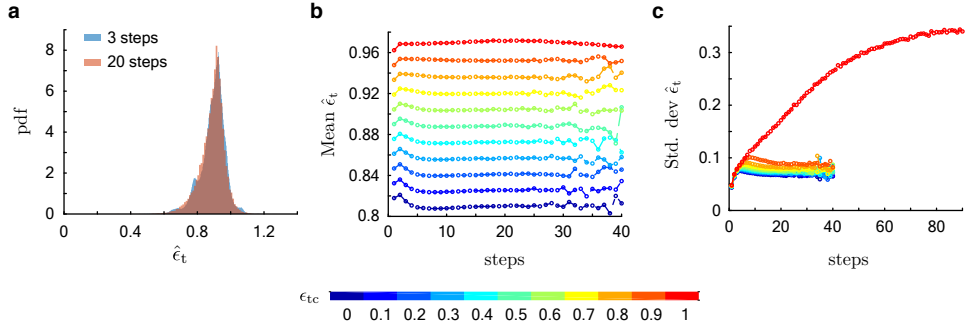

Figure S6: Evolution of forward collision parameter $\hat{\epsilon}_t$ as a function of step number and $\epsilon_{tc}$. **a**, The probability density function of $\hat{\epsilon}_t$ after 3 steps and 20 steps for $\epsilon_{tc} = 0.5$. **b**, Mean $\hat{\epsilon}_t$ as a function of step number for different values of $\epsilon_{tc}$ (shown in the color bar below). The mean converges by approximately 5 steps for all values of $\epsilon_{tc}$. **c**, Standard deviation of $\hat{\epsilon}_t$ as a function of step number for different values of $\epsilon_{tc}$. While the standard deviation converges by 15 steps for values of $\epsilon_{tc} < 1$, for $\epsilon_{tc} = 1$ it increases monotonically, plateauing only after approximately 80 steps.

5 steps (Fig. S6b), while the standard deviation for $\epsilon_{tc}$ sufficiently below 1 converges by approximately 15 steps (Fig. S6c). When $\epsilon_{tc} = 1$, or if $\epsilon_{tc}$ is sufficiently close to 1, the standard deviation of the distribution approaches steady state only after 80 steps.

## S5 Linear stability analysis: Jordan decomposition of $\boldsymbol{T}_{\text{ol}}$ and $\boldsymbol{T}_{\text{an}}$

The linear stability analysis for the open-loop runner proceeds analogously to the analysis for the anticipatory runner discussed in main text section 4. We define a Poincaré section transverse to the runner's trajectory in phase space at the apex of the aerial phase ($v_y = 0$), and a return map $\boldsymbol{f}_{\text{ol}}$ that maps the state of the runner $\boldsymbol{\psi}_n$ at the apex of the aerial phase at step $n$, to the state at the apex of the aerial phase on the following step $\boldsymbol{\psi}_{n+1}$ (main text Fig. 5). The state $\boldsymbol{\psi}$, return map $\boldsymbol{f}_{\text{ol}}$, and its linearization $\boldsymbol{T}_{\text{ol}}$ are defined by,

$$\boldsymbol{\psi} = \begin{pmatrix} x & y & \phi & v_x & \omega \end{pmatrix}^{\text{T}} \tag{S4a}$$

$$\boldsymbol{\psi}_{n+1} = \boldsymbol{f}_{\text{ol}}(\boldsymbol{\psi}_n), \tag{S4b}$$

$$\Delta\boldsymbol{\psi}_{n+1} = \boldsymbol{T}_{\text{ol}}\Delta\boldsymbol{\psi}_n, \text{ where } \boldsymbol{T}_{\text{ol}} = \left.\frac{\partial \boldsymbol{f}_{\text{ol}}}{\partial \boldsymbol{\psi}}\right|_{\boldsymbol{\psi}^*}, \tag{S4c}$$

$$\text{and } \Delta\boldsymbol{\psi} = \boldsymbol{\psi} - \boldsymbol{\psi}^*, \text{ where } \boldsymbol{\psi}^* = f_{\text{ol}}(\boldsymbol{\psi}^*). \tag{S4d}$$

The linearized return maps, $\boldsymbol{T}_{\text{ol}}$ and $\boldsymbol{T}_{\text{an}}$ are non-diagonalizable as discussed in main text section 4. Thus an eigen-factorization of these matrices is not possible, so we perform a Jordan decomposition

of $T_{\text{ol}}$ and $T_{\text{an}}$. The Jordan decomposition of $T_{\text{ol}}$ is,

$$T_{\text{ol}} = V_{\text{ol}} J_{\text{ol}} V_{\text{ol}}^{-1}, \tag{S5a}$$

$$\text{where } J_{\text{ol}} = \begin{pmatrix} 1 & 0 & \sqrt{2}\epsilon_{\text{n}} v_{y0} & 0 & 0 \\ 0 & 1 & \frac{\sqrt{2}}{1+I_{/\text{G}}} v_{y0}(I_{/\text{G}}(\epsilon_{\text{n}}-1)-\epsilon_{\text{n}}) & 0 & 0 \\ 0 & 0 & 1 & 0 & 0 \\ 0 & 0 & 0 & \epsilon_{\text{n}}(2\epsilon_{\text{n}}-1) & 0 \\ 0 & 0 & 0 & 0 & 0 \end{pmatrix}, \tag{S5b}$$

$$\text{and } V_{\text{ol}} = \begin{pmatrix} 0 & 1 & 0 & 0 & \frac{(1-I_{/\text{G}}-4\epsilon_{\text{n}})v_{y0}I_{/\text{G}}}{1+I_{/\text{G}}} \\ 0 & 0 & 0 & 1 & 0 \\ 1 & 0 & 0 & 0 & -v_{y0} \\ 0 & 0 & \frac{-1}{\sqrt{2}} & 0 & I_{/\text{G}} \\ 0 & 0 & \frac{1}{\sqrt{2}} & 0 & 1 \end{pmatrix}. \tag{S5c}$$

The linearized return map $T_{\text{ol}}$ has rank 4, with two stable eigenvalues; $\lambda = 0$ and $\lambda = \epsilon_{\text{n}}(2\epsilon_{\text{n}}-1)$. The second stable eigenvalue corresponds to perturbations to the height of the apex of the aerial phase. The remaining eigenvalues are $\lambda = 1$ which are associated with eigenvectors $\nu_1$, $\nu_2$, and generalized eigenvector $\nu_3$, the same as the eigenvectors and generalized eigenvector associated with eigenvalues $\lambda = 1$ for $T_{\text{an}}$ (main text equation (8a)). Thus, a perturbation in the subspace spanned by these vectors $\Delta\psi_0$ displays the same scaling with step number $n$ as described in main text equation (9) for the anticipatory runner. These relationships for the open-loop runner are summarized as follows,

$$\Delta\psi_0 = \sum_{k=1}^{3} \alpha_k \nu_k, \ \Delta\psi_n = n\alpha_3(a_1\nu_1 + a_2\nu_2) + \Delta\psi_0, \tag{S6a}$$

$$\Delta\psi_n \approx n\ \alpha_3 \begin{pmatrix} \frac{\sqrt{2}}{1+I_{/\text{G}}} v_{y0}(I_{/\text{G}}(\epsilon_{\text{n}}-1)-\epsilon_{\text{n}}) \\ 0 \\ \sqrt{2}\epsilon_{\text{n}} v_{y0} \\ 0 \\ 0 \end{pmatrix}, \text{ for } n \gg 1, \tag{S6b}$$

$$\text{substituting } a_1 = \sqrt{2}\epsilon_{\text{n}} v_{y0}, \text{ and } a_2 = \frac{\sqrt{2}}{1+I_{/\text{G}}} v_{y0}(I_{/\text{G}}(\epsilon_{\text{n}}-1)-\epsilon_{\text{n}}), \tag{S6c}$$

which are the off-diagonal terms of $J_{\text{ol}}$ (equation S5b). While the growth of the instability shows the same dependence on step number $n$ as the anticipatory runner (main text equation (9)), the value of $a_2$ differs, i.e. the projection onto $\nu_2$ of a perturbation along $\nu_3$ upon action of the return map $T_{\text{ol}}$.

The stability of the anticipatory runner is discussed in detail in main text section 4. The Jordan

decomposition of $\boldsymbol{T}_{\mathrm{an}}$ when $\epsilon_{\mathrm{tc}} < 1$ is,

$$\boldsymbol{T}_{\mathrm{an}} = \boldsymbol{V}_{\mathrm{an}} \boldsymbol{J}_{\mathrm{an}} \boldsymbol{V}_{\mathrm{an}}^{-1}, \tag{S7a}$$

$$\text{where } \boldsymbol{J}_{\mathrm{an}} = \begin{pmatrix} 1 & 0 & \sqrt{2}\epsilon_{\mathrm{n}} v_{y0} & 0 & 0 \\ 0 & 1 & -\sqrt{2}\epsilon_{\mathrm{n}} v_{y0} & 0 & 0 \\ 0 & 0 & 1 & 0 & 0 \\ 0 & 0 & 0 & \epsilon_{\mathrm{n}}(2\epsilon_{\mathrm{n}} - 1) & 0 \\ 0 & 0 & 0 & 0 & \epsilon_{\mathrm{t}} \end{pmatrix}, \tag{S7b}$$

$$\text{and } \boldsymbol{V}_{\mathrm{an}} = \begin{pmatrix} 0 & 1 & 0 & 0 & \frac{(1+\epsilon_{\mathrm{t}}(2\epsilon_{\mathrm{n}}-1))I_{/\mathrm{G}} v_{y0}}{\epsilon_{\mathrm{t}}-1} \\ 0 & 0 & 0 & 1 & 0 \\ 1 & 0 & 0 & 0 & \frac{(1+\epsilon_{\mathrm{t}}(2\epsilon_{\mathrm{n}}-1))v_{y0}}{\epsilon_{\mathrm{t}}-1} \\ 0 & 0 & \frac{-1}{\sqrt{2}} & 0 & I_{/\mathrm{G}} \\ 0 & 0 & \frac{1}{\sqrt{2}} & 0 & 1 \end{pmatrix}. \tag{S7c}$$

The anticipatory runner has a full rank return map $\boldsymbol{T}_{\mathrm{an}}$, and stable eigenvalues $\lambda = \epsilon_{\mathrm{n}}(2\epsilon_{\mathrm{n}} - 1)$, which is identical to the open-loop runner, and $\lambda = \epsilon_{\mathrm{t}}$ which differs from the open-loop runner. The off-diagonal elements of $\boldsymbol{J}_{\mathrm{an}}$ are $a_1 = \sqrt{2}\epsilon_{\mathrm{n}} v_{y0}$ and $a_2 = -\sqrt{2}\epsilon_{\mathrm{n}} v_{y0}$ (equation S7b), which correspond to the projection onto $\boldsymbol{\nu}_1$ and $\boldsymbol{\nu}_2$ respectively, of a perturbation along $\boldsymbol{\nu}_3$ upon action of the return map $\boldsymbol{T}_{\mathrm{an}}$.

When $\epsilon_{\mathrm{tc}} = \epsilon_{\mathrm{t}} = 1$ for the anticipatory runner, the linearization of $\boldsymbol{f}_{\mathrm{an}}$ about $\boldsymbol{\psi}^*$ (main text equations 7) yields a different form for $\boldsymbol{T}_{\mathrm{an}}$. The Jordan decomposition of $\boldsymbol{T}_{\mathrm{an}}$ when $\epsilon_{\mathrm{tc}} = 1$ is,

$$\boldsymbol{T}_{\mathrm{an}} = \boldsymbol{V}_{\mathrm{an}} \boldsymbol{J}_{\mathrm{an}} \boldsymbol{V}_{\mathrm{an}}^{-1}, \tag{S8a}$$

$$\text{where } \boldsymbol{J}_{\mathrm{an}} = \begin{pmatrix} 1 & 2\epsilon_{\mathrm{n}} v_{y0} & 0 & 0 & 0 \\ 0 & 1 & 0 & 0 & 0 \\ 0 & 0 & 1 & 2\epsilon_{\mathrm{n}} v_{y0} & 0 \\ 0 & 0 & 0 & 1 & 0 \\ 0 & 0 & 0 & 0 & \epsilon_{\mathrm{n}}(2\epsilon_{\mathrm{n}} - 1) \end{pmatrix}, \tag{S8b}$$

$$\text{and } \boldsymbol{V}_{\mathrm{an}} = \begin{pmatrix} 0 & 0 & 1 & 0 & 0 \\ 0 & 0 & 0 & 0 & 1 \\ 1 & 0 & 0 & 0 & 0 \\ 0 & 0 & 0 & 1 & 0 \\ 0 & 1 & 0 & 0 & 0 \end{pmatrix}. \tag{S8c}$$

There is a single stable eigenvalue of $\lambda = \epsilon_{\mathrm{n}}(2\epsilon_{\mathrm{n}} - 1)$ which is identical to that of the open-loop runner and the anticipatory runner with $\epsilon_{\mathrm{tc}} < 1$. The remaining 4 eigenvalues are $\lambda = 1$ which are associated with eigenvectors $\boldsymbol{\nu}_1$, $\boldsymbol{\nu}_2$ and generalized eigenvectors $\boldsymbol{\nu}_3$ and $\boldsymbol{\nu}_4$ (main text equations (10a)). Perturbations along $\boldsymbol{\nu}_3$ project onto $\boldsymbol{\nu}_1$, and perturbations along $\boldsymbol{\nu}_4$ project onto $\boldsymbol{\nu}_2$, upon action of the return map $\boldsymbol{T}_{\mathrm{an}}$ (equation (S8b)). The off-diagonal elements of $\boldsymbol{J}_{\mathrm{an}}$ are $a_1 = a_2 = 2\epsilon_{\mathrm{n}} v_{y0}$.

## S6 Scaling analysis of orientational failures

In main text section 5, we show that the orientation failure bound $\phi_{\mathrm{tol}}$ and the orientation change over a single step $\phi_{\bullet}$ from encountering an unexpected terrain slope predict the mean steps to failure $N$, as $N \sim \phi_{\mathrm{tol}}/\phi_{\bullet}$. The many parameters that define the runner and its gait do not separately

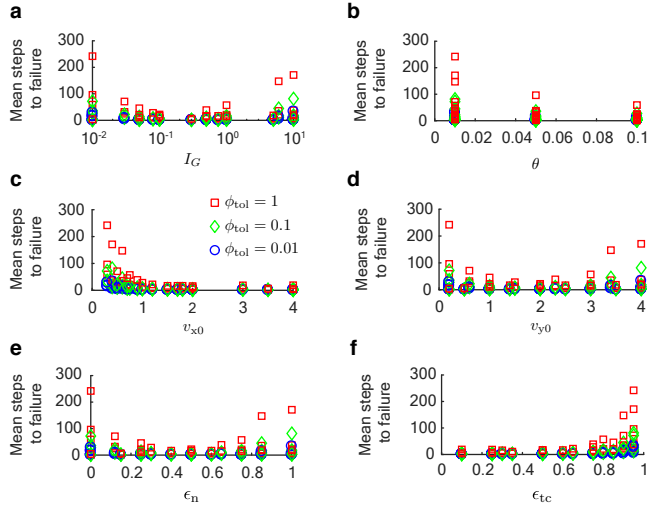

Figure S7: Mean steps to failure for the anticipatory runner ($\epsilon_{tc} < 1$) plotted against all other parameters: $I_{/G}, \theta, v_{y0}, \epsilon_n, v_{x0}, \epsilon_{tc}$. Here, $\theta$ refers to the typical slope angle the runner would encounter, defined as $\theta = h/\lambda$, where $\lambda$ is the grid spacing of the terrain and $[-h, h]$ is the range of heights of the terrain (main text section 7.3).

predict mean steps to failure as accurately as the parameter $\phi_{tol}/\phi_\bullet$ (Fig. S7).

The change in orientation over one step can be calculated using the vertical velocity $v_{y,\bullet}^+$ and angular velocity $\omega_\bullet^+$ at take-off as $\phi_\bullet = 2v_{y,\bullet}^+ \omega_\bullet^+$ ($\bullet$ = 'ol' for open-loop and 'an' for anticipatory). For a runner landing on a terrain patch angled at $\theta$ with respect to the horizontal, with center of mass velocity $\boldsymbol{v}_G^- = \begin{pmatrix} v_{x0} & -v_{y0} \end{pmatrix}^T$ and angular velocity $\omega^- = 0$, the take-off velocities $v_{y,\bullet}^+, \omega_\bullet^+$, and orientation change $\phi_\bullet$, are given by

$$\omega_{an}^+ = \frac{(1-\epsilon_{tc})(v_{x0}(1-\cos\theta) + v_{y0}\sin\theta)}{1+I_{/G}}, \tag{S9a}$$

$$\omega_{ol}^+ = \frac{v_{x0}(1-\cos\theta) + v_{y0}\sin\theta}{1+I_{/G}}, \tag{S9b}$$

$$v_{y,an}^+ = v_{y0} + \frac{(1+\epsilon_n + I_{/G}\epsilon_n)(v_{x0}\sin 2\theta - v_{y0}(1-\cos 2\theta))}{2(1+I_{/G})}, \tag{S9c}$$

$$v_{y,ol}^+ = v_{y,an} + \frac{2\epsilon_{tc}v_{x0}\sin\theta}{2(1+I_{/G})}, \tag{S9d}$$

$$\phi_{an} = \frac{1-\epsilon_{tc}}{(1+I_{/G})^2} A\big(B + (\epsilon_{tc}-1)C + (D+I_{/G}\epsilon_{tc})E + F - I_{/G}\epsilon_{tc}v_{y0}\big),$$

$$\phi_{ol} = \frac{1}{(1+I_{/G})^2} A\big(B + (2\epsilon_{tc}-1)C + DE + F\big), \tag{S9e}$$

where $A = v_{y0}\sin\theta + v_{x0}(1-\cos\theta)$, $B = 2(1+I_{/G})(1-\epsilon_n)v_{y0}\cos\theta$,
$C = 2I_{/G}v_{x0}\sin\theta$, $D = 1+\epsilon_n + I_{/G}\epsilon_n$,
$E = v_{y0}\cos 2\theta + v_{x0}\sin 2\theta$, $F = (I_{/G}\epsilon_n + \epsilon_n - 1)v_{y0}$.

The parametric dependence of $\phi_{tol}/\phi_\bullet$ ($\phi_\bullet$ defined in equation (S9e)) on $\epsilon_{tc}$ and $\epsilon_n$ is captured by a power series approximation of $\phi_\bullet$ to second order in $\theta$ (main text equations (12a)-(12b)). The

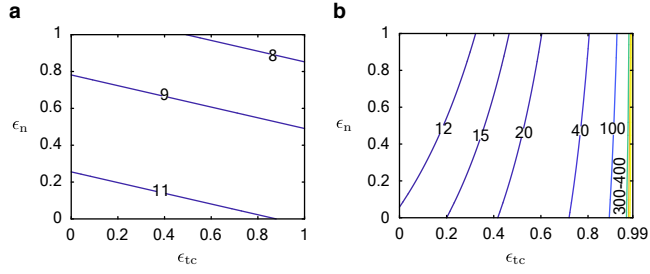

Figure S8: Contour plots of **a**, $\phi_{\text{tol}}/\phi_{\text{ol}}$ and **b**, $\phi_{\text{tol}}/\phi_{\text{an}}$ as a function of $\epsilon_n$ and $\epsilon_{\text{tc}}$, where the form of $\phi_\bullet$ shown in the power series expansion to second order in $\theta$ has been used (main text equations (12a)-(12b)). The values of the contours differ slightly between the plots shown here and in the main text Fig. 7 but the qualitative trends are the same. The same parameter values are used in both these calculation, $v_{y0} = 0.2, I_{/\text{G}} = 0.15, v_{x0} = 1, \theta = 0.3, \phi_{\text{tol}} = 1$. For the anticipatory runner, maximum $\epsilon_{\text{tc}} = 0.99$ to ensure that the calculation is limited to the region where runners primarily undergo orientational failures (Fig. S5a).

contour plots of $\phi_{\text{tol}}/\phi_\bullet$ shown here using the truncated power series form of $\phi_\bullet$ (Fig. S8) display the same qualitative trends as the ones in main text Fig. 7 which use the complete expression for $\phi_\bullet$ (equation (S9e)). Therefore, we analyze failure statistics and runner morphology in main text section 5 and section 6 using the truncated power series approximation for $\phi_\bullet$.

## S7 Steps to failure statistics

All steps to failure distributions show similar qualitative features; they are unimodal with a long tail, and are relatively insensitive to changes in the terrain height distribution (Fig. S1, main text Fig. 2e). This suggests that the stochastic dynamics underlying these distributions are characterized by a small number of parameters that govern angular momentum fluctuations of the runner, as most runners fail by losing orientational stability.

To arrive at this lower-order model, we start with the result from the linear stability analysis of the runner which states that the orientation of the runner $\phi$ grows linearly with step number $n$ in response to small perturbations to the periodic orbit (main text section 4). A second-order difference equation that displays linear growth in step number $n$ of an initial perturbation $\phi(1) - \phi(0)$ is,

$$\phi(n+2) = 2\phi(n+1) - \phi(n), \tag{S10a}$$
$$\implies \phi(n) = n(\phi(1) - \phi(0)) + \phi(0). \tag{S10b}$$

For the runner, perturbations arise from terrain slope variations and are thus stochastic in nature. Thus, we incorporate stochasticity in this model using an additive noise term,

$$\phi(n+2) = 2\phi(n+1) - \phi(n) + w, \tag{S11a}$$
$$\text{where } w \sim N(0, \sigma^2), \tag{S11b}$$

and $N(0, \sigma^2)$ is the normal distribution with mean $= 0$ and standard deviation $= \sigma$. We assume that after a sufficiently large number of steps, $\phi(n)$ would have Gaussian statistics as terrain perturbations are uncorrelated from step-to-step. This invocation of the mean value theorem, while not rigorous, is done primarily to simplify the analysis. With initial conditions $\phi(1) = \phi(0) = 0$,

the probability distribution of $\phi$ after $n$ steps is given by

$$p(\phi, n) = \frac{e^{-\frac{\phi^2}{2\sigma^2 f(n)}}}{\sqrt{2\pi\sigma^2 f(n)}}, \tag{S12a}$$

$$\text{where } f(n) = \frac{n(n+1)(2n+1)}{6} \tag{S12b}$$

$$\text{and } f(n) \sim \frac{n^3}{3}, \text{ for } n \gg 1. \tag{S12c}$$

The probability distribution $p(\phi, n)$ is Gaussian with $<\phi> = 0$ and variance that grows as $n^3$ for large $n$.

Orientation failures occur when a runner's orientation exceeds $\phi_{\text{tol}}$. Thus, the probability of having failed by $n$ steps, which we denote with $p_{\text{fall}}(n)$, is given by

$$p_{\text{fall}}(n) = p(|\phi| > \phi_{\text{tol}}, n), \tag{S13a}$$

$$\implies p_{\text{fall}}(n) = 1 - \text{erf}\left(\frac{\phi_{\text{tol}}}{\sqrt{2\sigma^2 f(n)}}\right), \tag{S13b}$$

$$\text{where } \text{erf}(x) = \frac{1}{\sqrt{\pi}} \int_{-x}^{x} e^{-u^2} du, \tag{S13c}$$

is the error function. The probability of failing at step number $n$, denoted by $p_{\text{steps}}(n)$, is given by the probability of taking $n-1$ steps without failing, and then failing at step number $n$, i.e. $p_{\text{steps}}(n) = p(|\phi| > \phi_{\text{tol}}, n \big| |\phi| < \phi_{\text{tol}}, n-1)$. This is equivalent to the probability flux at $\phi_{\text{tol}}$ at step $n$, given by

$$p_{\text{steps}}(n) = p_{\text{fall}}(n) - p_{\text{fall}}(n-1), \tag{S14a}$$

$$\implies p_{\text{steps}}(n) = \text{erf}\left(\frac{\phi_{\text{tol}}}{\sqrt{2\sigma^2 f(n-1)}}\right) - \text{erf}\left(\frac{\phi_{\text{tol}}}{\sqrt{2\sigma^2 f(n)}}\right). \tag{S14b}$$

The mean steps to failure $M$ are thus given by

$$M = \sum_{n=0}^{\infty} n p_{\text{steps}}(n), \tag{S15a}$$

$$\implies M = \sqrt{\frac{6}{\pi}} \sum_{i=1}^{\infty} -1^{i+1} \left(\frac{3}{2}\right)^i \zeta\left(\frac{3(2i-1)}{2}\right) \frac{(\phi_{\text{tol}}/\sigma)^{2i-1}}{(i-1)!}, \tag{S15b}$$

$$\text{where we have used } f(n) \approx \frac{n^3}{3}, \tag{S15c}$$

and $\zeta(p)$ is the Riemann-zeta function defined by $\zeta(p) = \sum_{i=0}^{\infty} 1/i^p$ for $p > 1$.

The probability of failing at step $n$, $p_{\text{steps}}(n)$ (Fig. S9a), and it's cumulative distribution $p_{\text{fall}}(n)$ (Fig. S9b) are qualitatively similar to the numerically derived steps to failure distributions shown in Fig. S1 and main text Fig. 2e, and the cumulative distributions shown in main text Fig. 2a,b.

Interpreting the standard deviation of the noise from the terrain as $\sigma = \phi_\bullet$, since $\phi_\bullet$ sets the scale for orientation change over a single step due to an unexpected terrain slope (main text section 5), we do not find quantitative agreement between our stochastic model and the numerics. The mean

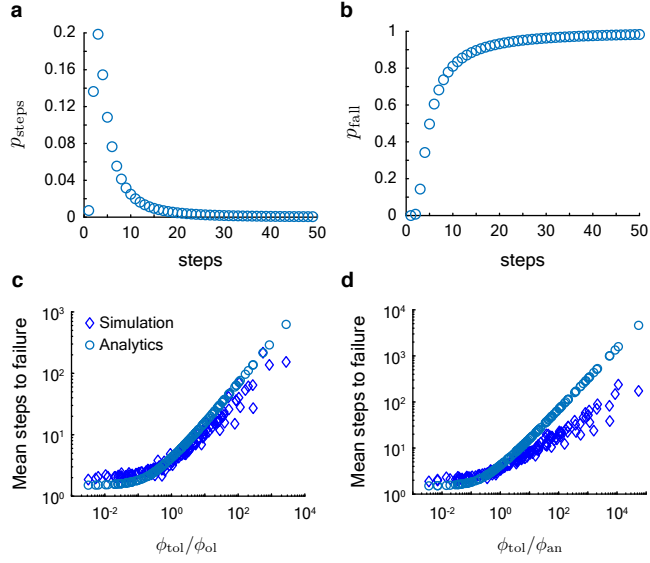

Figure S9: Steps to failure statistics of the single parameter stochastic model. **a**, $p_{\text{steps}}$ from equation (S14) and **b**, $p_{\text{fall}}$ from equation (S13) plotted against steps taken $n$ with human-like parameter values for $\phi_{\text{tol}}$ and $\sigma = \phi_{\text{ol}}$ (main text table 1). The mean of this distribution 9.6, is remarkably similar to the numerical mean steps to failure for the open-loop runner with human-like parameters which is $9.61 \pm 0.01$. However, for a wider range of $\phi_{\text{tol}}/\phi_{\bullet}$ values, the mean of the analytical model (open circles) for **c**, open-loop runners and **d**, anticipatory runners deviates from mean steps to failure from the Monte Carlo simulations (open diamonds, main text Fig. 6). The parameters values used for $\phi_{\text{tol}}$ and $\phi_{\bullet}$ are shown in Fig. S7.

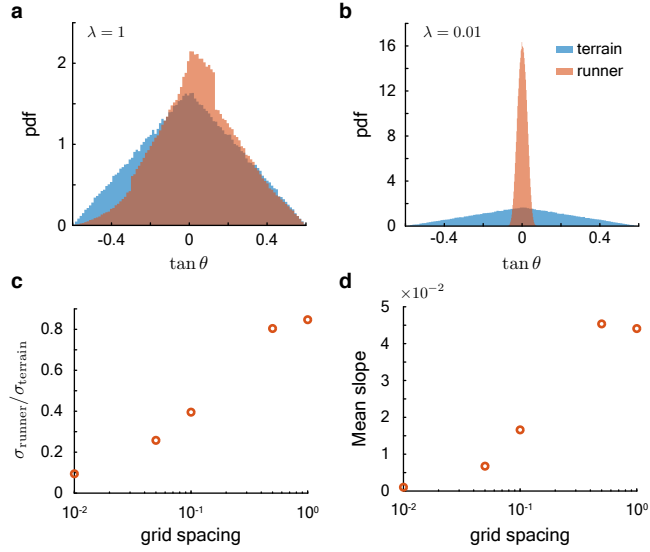

Figure S10: Effect of varying the ratio of leg length $r_\ell$ to terrain grid spacing $\lambda$. The probability density function of the terrain slope ($\tan\theta$) encountered by the open-loop runner (orange) with human-like parameters is plotted for grid spacing **a**, $\lambda = r_\ell$ and **b**, $\lambda = 0.01 r_\ell$. The height distribution of the terrain grid points was adjusted so that the terrain's slope distribution (blue) was the same in all simulations. The number of steps used to generate these probability density functions was $\sim 10^6$. **c**, Shows the standard deviation of the slope encountered by the runner relative to the standard deviation of the terrain's slope distribution. **d**, The mean slope encountered by the runner is always positive and increases with grid spacing, even though the terrain's slope distribution is held constant.

of the analytical distribution (equation (S15b)) has a different power law scaling compared to mean steps to failure obtained from the Monte Carlo simulations (Fig. S9c,d).

## S8 Effect of changing runner size relative to terrain grid spacing

The size of the runner $r_\ell$, relative to the grid spacing of the terrain $\lambda$, determines the range of terrain slopes accessed by the runner (main text section 6, Fig. S10). For example, if $\lambda/r_\ell = 1$, the range of terrain slopes encountered by the runner is nearly the same as the slope distribution of the terrain itself (Fig. S10a). By decreasing this ratio to $\lambda/r_\ell = 0.01$ while keeping the terrain slope distribution constant, we find that the range of slopes encountered by the runner is significantly lower (Fig. S10b). We quantify this trend using the parameter $\sigma_{\text{runner}}/\sigma_{\text{terrain}}$ where $\sigma_{\text{terrain}}$ is the standard deviation of the terrain slope distribution, and $\sigma_{\text{runner}}$ is the standard deviation of the slope distribution encountered by the runner. The parameter $\sigma_{\text{runner}}/\sigma_{\text{terrain}}$ approaches 1 as $\lambda/r_\ell$ increases (Fig. S10c).

The mean slope encountered by the runner is always greater than zero (Fig. S10d) even though the slope distribution of the terrain has a zero mean. As the range of slopes encountered by the runner increases, the mean slope encountered by the runner deviates further from zero (Fig. S10d). This is consistent with the observation from the Monte Carlo simulations that runners slow down on rough terrain since the effect of the mean slope encountered is to redirect forward momentum into vertical momentum.

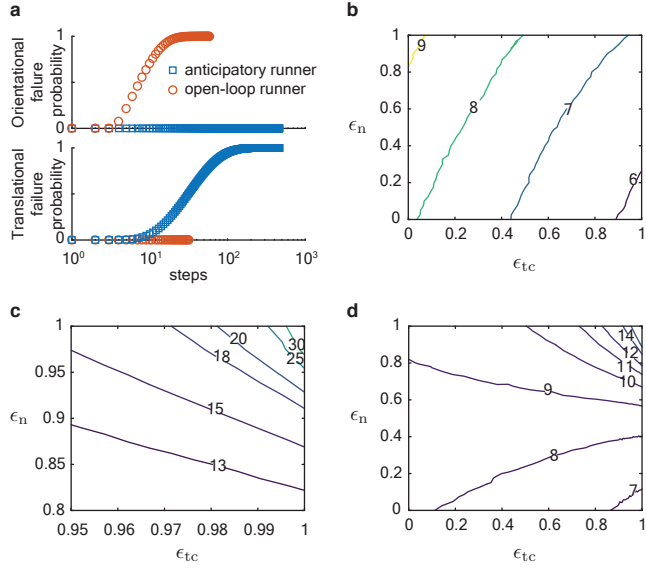

Figure S11: Runners with a *lab-fixed* push-off policy: effect of energy dissipation and tangential collisions. **a**, Open-loop runners with human-like parameters and $\epsilon_{tc} = 0, \epsilon_n = 1$ undergo orientational failures on rough terrain while human-like anticipatory runners with $\epsilon_{tc} = 1, \epsilon_n = 1$ maintain orientation but lose forward momentum. **b**, Contour map of mean steps to failure for the human-like open-loop runner as a function of $\epsilon_n$ and $\epsilon_{tc}$. **c**, Contour map of the mean steps to failure for the anticipatory runner as a function of $\epsilon_n$ and $\epsilon_{tc}$, zoomed into the region near the maximum steps taken, where contour spacing is small. **d**, Away from the maximum, contour spacing increases for the anticipatory runner, demonstrating reduced sensitivity to changes in $\epsilon_n$ and $\epsilon_{tc}$. The $\epsilon_{tc} = 0$ strategy is identical between the open-loop and anticipatory runners like with the *terrain-fixed* push-off policy.

## S9 Monte Carlo simulations with a *lab-fixed* push-off policy

The change in angular velocity due to the push-off impulse $\boldsymbol{J}_{imp}$ applied under the *lab-fixed* push-off policy is terrain dependent and destabilizes the runner as it exacerbates angular momentum fluctuations that arise from variations in terrain slope angle $\theta$. When $\theta > 0$, the runner collides with lower tangential velocity as compared to landing on flat ground, yet the change in angular velocity at push-off $\omega_{imp}$, is larger than the corresponding impulse on flat ground (equation S2d) causing the runner to spin excessively in the anti-clockwise direction. Conversely, when $\theta < 0$, the runner lands at a shallower angle, increasing the tangential collision impulse. The change in angular velocity $\omega_{imp}$ is lower than on flat ground (equation S2d), causing the runner to spin in the clockwise direction. In contrast, $\omega_{imp}$ under the *terrain-fixed* push-off policy is unchanged by terrain slope $\theta$ (equation S2d) and thus, in all cases, the *lab-fixed* push-off policy leads to a lower number of steps taken compared to the *terrain-fixed* push-off policy.

### S9.1 Open-loop runners

Increasing energy dissipation in the direction normal to the terrain surface reduces mean steps taken by open-loop runners with a *lab-fixed* push-off policy in contrast to the trend observed with the *terrain-fixed* push-off where increasing normal energy dissipation, nominally increased steps taken (main text Fig. 2d). With the *lab-fixed* push-off, open-loop runners with human-like parameters

S15

only fail due to losing orientational stability (Fig. S11a) and increase mean steps taken by at most 1 for a 100% reduction in energy dissipation from $\epsilon_n = 0$ to $\epsilon_n = 1$ at a fixed value of $\epsilon_{tc}$ (Fig. S11b).

Reducing tangential collisions increases steps taken (Fig. S11b), in contrast to the trend observed with the *terrain-fixed* push-off strategy where changing $\epsilon_{tc}$ had negligible effect on the mean steps to failure (main text Fig. 2d). Open-loop runners increase mean steps taken by 2 when $\epsilon_{tc}$ is reduced from 1 to 0, while holding $\epsilon_n$ fixed at any value (Fig. S11b).

Open-loop runners with a *lab-fixed* push-off policy perform best when $\epsilon_n = 1, \epsilon_{tc} = 0$ and worst when $\epsilon_n = 0, \epsilon_{tc} = 1$. The difference in mean steps taken at these extremes is just 3 steps, with the maximum of 9 steps being the same as the minimum number of steps taken by human-like open-loop runners with the *terrain-fixed* push-off policy (main text Fig. 2d). Thus, in contrast to the *terrain-fixed* push-off policy, open-loop runners with the *lab-fixed* push-off policy fare worse, and display differing trends in how stability is determined by the collision parameters. However, similar to the *terrain-fixed* push-off policy, varying parameters governing the passive collision, $\epsilon_n$ and $\epsilon_{tc}$, results in only a small effect on the mean steps to failure compared to anticipatory runners.

### S9.2 Anticipatory runners

Anticipatory runners with a *lab-fixed* push-off policy increase steps taken as energy dissipation in the direction normal to the terrain surface is reduced, similar to open-loop runners with the same push-off policy. Like for the open-loop runners, this is in contrast to the trend observed for anticipatory runners with the *terrain-fixed* push-off policy, where increasing normal energy dissipation increases mean steps to failure (main text Fig. 3a). Anticipatory runners fail by losing orientational stability for a vast majority of values of the collision parameters, and orientation is maintained only when $\epsilon_{tc} = 1$ and $\epsilon_n = 1$ (Fig. S11a) in contrast to the *terrain-fixed* push-off policy where anticipatory runners maintain orientation when $\epsilon_{tc} = 1$, and regardless of the value of $\epsilon_n$ (main text Fig. 3a). At $\epsilon_{tc} = 1$, anticipatory runners with a *lab-fixed* push-off, increase mean steps taken by over 4-fold, from 7 to over 30 for a 100% reduction in normal energy dissipation from $\epsilon_n = 0$ to $\epsilon_n = 1$ (Fig. S11c).

Reducing the tangential collisional impulse at landing increases mean steps to failure for the anticipatory runner with a *lab-fixed* push-off if $\epsilon_n$ is sufficiently high, while showing the opposite trend below that $\epsilon_n$ value (Fig. S11d). This is in contrast to the trend observed in anticipatory runners with the *terrain-fixed* push-off policy where mean steps taken always increases with reduced normal energy dissipation (main text Fig. 3a).

Anticipatory runners with a *lab-fixed* push-off policy perform best when $\epsilon_{tc} = 1, \epsilon_n = 0$ and worst when $\epsilon_{tc} = 1, \epsilon_n = 0$. The *lab-fixed* push-off policy is consistently worse in terms of stability than the *terrain-fixed* push-off policy for a given pair of $\epsilon_{tc}, \epsilon_n$ values except for $\epsilon_{tc} = 1, \epsilon_n = 1$ where both push-off policies are identical, as this corresponds to $\boldsymbol{J}_{imp} = 0$ in both.

Running with non-zero tangential collisions is optimal in the presence of sensorimotor noise (Fig. S12a). Sensorimotor noise is modeled as additive noise to $\epsilon_{tc}$ as described in main text section 3.5. This trend is similar to that found with the *terrain-fixed* push-off policy (main text Fig. 3b).

### S9.3 Scaling analysis of mean steps taken

The mean statistics for the steps to orientational failure on rough terrains are predicted by the orientation change over a single step relative to the failure bound $\phi_{tol}/\phi_\bullet$ (Fig. S12c,d). The loss of forward speed relative to initial forward speed $(v_{x0}/\Delta v_x)$ predicts failure statistics due to losing translational stability for anticipatory runners with $\epsilon_{tc} = 1, \epsilon_n = 1$ (Fig. S12b). These trends are

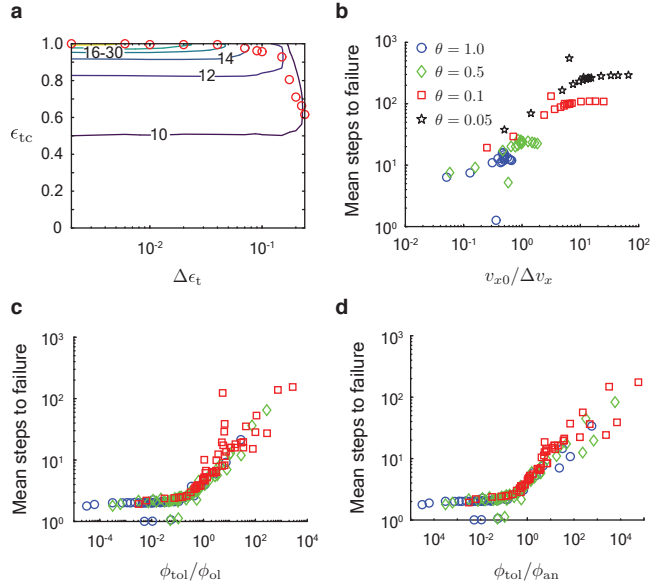

Figure S12: Effect of sensorimotor noise on stability, and scaling analysis of mean steps to failure. **a**, Contour map of mean steps taken as a function of $\epsilon_{\text{tc}}$ and $\Delta\epsilon_{\text{t}}$ shows that mean steps to failure reduces with increasing sensorimotor noise $\Delta\epsilon_{\text{t}}$. Scuffing the ground upon landing is optimal as noise intensity $\Delta\epsilon_{\text{t}}$ is increased. Red circles denote the optimal $\epsilon_{\text{tc}}$ value for a given $\Delta\epsilon_{\text{t}}$. In these simulations $\epsilon_{\text{n}} = 1$. **b**, For the anticipatory runner with $\epsilon_{\text{tc}} = 1, \epsilon_{\text{n}} = 1$, the mean steps to translational failure are predicted by the parameter $v_{x0}/\Delta v_x$. The values of the remaining parameters used in this simulation are shown in Fig. S7. Mean steps to orientation failure are captured by the parameter **c** $\phi_{\text{tol}}/\phi_{\text{ol}}$ for the open-loop runner, and by **d** $\phi_{\text{tol}}/\phi_{\text{an}}$, for the anticipatory runner. The parameter values used for these simulations are shown in Fig. S7.

similar to those observed for the runner with a *terrain-fixed* push-off policy (main text Fig. 6), although there is a greater spread in the data about the trend line with the *lab-fixed* push-off.



\end{document}